\documentclass{pasj00} \draft
\begin{document}
\SetRunningHead{M.Shimizu et al.}{Non-Gravitational Heating Processes
  in Galaxy Clusters}   \Received{2003 September 23} \Accepted{2003
  October 30}

\title{Monte-Carlo Modeling of Non-Gravitational Heating Processes in
  Galaxy Clusters}

\author{Mamoru \textsc{Shimizu},\altaffilmark{1}
  
  Tetsu \textsc{Kitayama},\altaffilmark{2}
  
  Shin \textsc{Sasaki},\altaffilmark{3}
  
  and Yasushi \textsc{Suto}\altaffilmark{1}}

\altaffiltext{1}{Department of Physics, School of Science, The
  University of Tokyo, Tokyo 113-0033}

\altaffiltext{2}{Department of Physics, Toho University,  Funabashi,
  Chiba 274-8510}

\altaffiltext{3}{Department of Physics, Tokyo Metropolitan University,
  Hachioji, Tokyo 192-0397}

\email{mshimizu@utap.phys.s.u-tokyo.ac.jp, kitayama@ph.sci.toho-u.ac.jp,\\
  sasaki@phys.metro-u.ac.jp, suto@phys.s.u-tokyo.ac.jp}


\maketitle

\begin{abstract}
  We consider non-gravitational heating effects on galaxy clusters on
  the basis of the Monte-Carlo modeling of merging trees of dark
  matter halos combined with the thermal evolution of gas inside each
  halo.  Under the assumption of hydrostatic equilibrium and the
  isothermal gas profiles, our model takes account of the metallicity
  evolution, metallicity-dependent cooling of gas,  supernova energy
  feedback, and heating due to jets of radio galaxies in a consistent
  manner.  The observed properties of galaxy clusters can be explained
  in models with higher non-gravitational heating efficiency than that
  in the conventional model. Possibilities include jet heating by the
  Fanaroff-Riley Type II radio galaxies, and the enhanced star
  formation efficiency and/or supernova energy feedback, especially at
  high redshifts.
\end{abstract}

%%%%%%%%%%%%%%%%%%%%%%%%%%%%%%%%%%%%%%%%%%%%%%%%%%%%%%%%%%%%%%%%%%%%%%
\section{Introduction}
%%%%%%%%%%%%%%%%%%%%%%%%%%%%%%%%%%%%%%%%%%%%%%%%%%%%%%%%%%%%%%%%%%%%%%

Energy feedback plays a vital role in a variety of different phenomena
and scales of the universe. Supernova explosion produces an overdense
shock shell in the surrounding interstellar medium which triggers the
subsequent star formation. The Gunn-Peterson test in the quasar
spectra has revealed that our universe was reionized at high redshifts
($z \sim 6$).  The recent \textit{WMAP} result indeed indicated that
the reionization epoch may be even earlier than previously thought, $z
= 17 \pm 5$ \citep{Spergel03}. This implies that the energy feedback
from first cosmological objects, or \textit{non-gravitational heating
  of the universe}, was much stronger than the conventional model
predictions.

Non-gravitational heating is also believed to have had a significant
influence on the scales of galaxy clusters.  This is clearly
illustrated by the well-known inconsistency of the observed X-ray
luminosity-temperature ($L_{\mathrm{X}}$-$T$) relation;
$L_{\mathrm{X}} \propto T^3$ (e.g., \cite{david93};
\cite{markevitch98}; \cite{ae99}) against the simple self-similar
prediction $L_{\mathrm{X}} \propto T^2$ \citep{kaiser86}.  More
recently \citet{vb01} suggested that the effect of cooling is
important in reproducing the observed $L_{\mathrm{X}}$-$T$ relation
(see also \cite{wx02}). The subsequent simulations (e.g.,
\cite{muan02, kay03, tornatore03}), however, indicate that while the
$L_{\mathrm{X}}$-$T$ relation can be explained by purely cooling
effect, the observed hot gas fraction requires a fairly significant
amount of non-gravitation heating (e.g., figure 3 of \cite{muan02} and
figure 8 of \cite{kay03}).

Physical origin of the non-gravitational heating of the intracluster
medium (ICM) still remains to be understood. A plausible candidate
responsible for the heating is the energy feedback by supernova
explosions before and/or during the formation of galaxy clusters
\citep{eh91, kaiser91}.  Previous authors (\cite{cavaliere99};
\cite{balogh99}; \cite{pen99}; \cite{ky00}; \cite{loewenstein00};
\cite{wu00}; \cite{bower01}; \cite{bm01}), however, concluded that the
excess heating energy $\sim 1$keV per gas particle is required to
account for the observed $L_{\mathrm{X}}$-$T$ relation.  This amount
of energy seems larger than the conventional model prediction of the
supernova combined with the standard star formation history and the
initial mass function of stars.  Another candidate of the heating
sources frequently discussed is active galactic nuclei (AGNs). Since
the radiation from AGNs is ineffective in heating the ICM, one has to
look for the kinetic energy input, from radio jets for instance, to
efficiently thermalize the ICM \citep{wu00}. \citet{IS01} claimed that
jets from radio galaxies can provide sufficient energy input to
thermalize the ICM and explain the observed $L_{\mathrm{X}}$-$T$
relation on the basis of simple analytic estimates, despite several
considerable uncertainties of the abundances and intrinsic properties
of such radio galaxies.

In this paper, we explore extensively the consequences of
non-gravitational heating processes on the observable properties of
galaxy clusters. For that purpose, we follow the merging history of
dark matter halos in the Monte-Carlo fashion and then trace the
thermal evolution of baryonic gas inside these halos. This approach
was first applied by \citet{wu00} in the context of the ICM heating,
and our current method improves their modeling in several aspects and
also approaches the problem in a complementary fashion; (i) we follow
the merging tree of dark halos of mass down to $1.7 \times 10^{7}
M_\odot$ from $z=30$ to $z=0$ so as to resolve all the halos that may
cool, i.e., whose virial temperature exceeds $10^4$K
\citep{mshimizu02}.  (ii) We simultaneously consider the supernova
energy and the jets of radio galaxies as heating sources in addition
to various cooling processes. Thus the thermal evolution of baryonic
gas, star formation history and the metallicity evolution are solved
in a consistent manner. (iii) We adopt the cooling rate which
incorporates the metallicity evolution of hot gas in each halo.  (iv)
The overall strength of heating in our model is controlled by the two
dimensionless parameters, $\epsilon_{\scriptscriptstyle\mathrm{SN}}$
and $\epsilon_{\scriptscriptstyle\mathrm{RG}}$.  Their values are
normalized so that $\epsilon_{\scriptscriptstyle\mathrm{SN}} =
\epsilon_{\scriptscriptstyle\mathrm{RG}}=1$ for our canonical sets of
assumptions. Nevertheless we survey a wider range of the parameter
space taking account of the fact that the nature of those sources are
poorly understood, especially at high redshifts. (v) We also consider
models with the enhanced star formation efficiency at high redshifts
($z>7$) so as to look for any possible implications on the cluster
$L_{\mathrm{X}}$-$T$ relation on the basis of the recent \textit{WMAP}
suggestion of the early reionization in the universe \citep{Spergel03}.

The rest of the paper is organized as follows.  Section 2 briefly
describes our method of tracing merger trees of dark matter halos.
The basic picture of the non-gravitational heating processes,
supernova feedback and jets of radio galaxies, is presented in section
3.  Our model to follow thermal and metallicity evolution of baryonic
gas inside dark halos is shown in section 4. We derive constraints on
our model parameters from the observed metallicity -- temperature
relation (\S 5) and $L_{\mathrm{X}}$-$T$ relation (\S 6).  Section 7
presents further comparison between our predictions and the observed
properties of galaxy clusters in X-ray band. We also briefly compare
our results with that of \citet{wu00}.  We consider the enhanced star
formation model at high redshift in section 8.  Finally section 9 is
devoted to summary and conclusions.

Throughout the paper, we adopt a conventional $\Lambda$CDM model with
the following set of cosmological parameters (e.g., \cite{Spergel03});
the density parameter $\Omega_{\mathrm{M}}=0.3$, the cosmological
constant $\Omega_{\Lambda}=0.7$, the dimensionless Hubble constant
$h_{70}\equiv H_{0}/(70 \;\mathrm{km}\;\mathrm{s}^{-1} \;
\mathrm{Mpc}^{-1})=1$, the baryon density parameter
$\Omega_{\mathrm{B}}=0.04h_{70}^{-2}$, and the value of the mass
fluctuation amplitude at $8\;h^{-1}\;\mathrm{Mpc}$, $\sigma_8=0.84$,
where $h=0.7h_{70}$.

%%%%%%%%%%%%%%%%%%%%%%%%%%%%%%%%%%%%%%%%%%%%%%%%%%%%%%%%%%%%%%%%%%
\section{Modeling Merger Trees of Dark Matter Halos} \label{sec:tree}
%%%%%%%%%%%%%%%%%%%%%%%%%%%%%%%%%%%%%%%%%%%%%%%%%%%%%%%%%%%%%%%%%%

Our approach begins with constructing realizations of merger histories
of dark matter halos.  Specifically we use the method
\citep{mshimizu02} that is based on an improved version of the
algorithm first proposed by \citet{SK99}.

Consider a halo of mass $M_1$ located at redshift $z_1$.  The mass of
its progenitors, $M_2$, at a slightly earlier redshift $z_2=z_1+\Delta
z(z_1)$ obeys the \textit{mass-weighted} conditional probability
function derived in the extended Press-Schechter theory
\citep{bower91,bond91}:
%%%%%%%%%%%%%%%%%%%%%%%%%%%%%%%%%%%%%%%%%%%%%%%%%%%%%%%%%%%%%%%%%%%
\begin{equation}
  \label{eq:eps-mass}
  \frac{dP}{dM_{2}}(M_{2},z_{2}|M_{1},z_{1})
  =\frac{\delta_{\mathrm{c},2}-\delta_{\mathrm{c},1}}
  {\sqrt{2\pi(S_{2}-S_{1})^{3}}}\
  \exp\!\left(
    - \frac{(\delta_{\mathrm{c},2}-\delta_{\mathrm{c},1})^{2}}
    {2(S_{2}-S_{1})} 
  \right)
  \left|
    \frac{dS_{2}}{dM_{2}}
  \right| ,
\end{equation}
%%%%%%%%%%%%%%%%%%%%%%%%%%%%%%%%%%%%%%%%%%%%%%%%%%%%%%%%%%%%%%%%%%%
where $\delta_{\mathrm{c},i} \sim 3 (12\pi)^{2/3} /20 D(z_i)$ (its
useful approximate formula may be found in Kitayama, Suto 1996) is the
critical over-density of the mass density field at a redshift of
$z_{i}$, $D(z_i)$ is the linear growth rate, and
$S_{i}\equiv\sigma^{2}(M_{i})$ is a mass variance of the density field
top-hat smoothed over the mass scale $M_i$.  The corresponding
\textit{number-weighted} conditional probability function for $M_2$ is
written as
%%%%%%%%%%%%%%%%%%%%%%%%%%%%%%%%%%%%%%%%%%%%%%%%%%%%%%%%%%%%%%%%%%%
\begin{equation}
  \label{eq:eps-num}
  \frac{dN}{dM_{2}}(M_{2},z_{2}|M_{1},z_{1})
  =\frac{M_{1}}{M_{2}}\frac{dP}{dM_{2}}(M_{2},z_{2}|M_{1},z_{1}) .
\end{equation}
%%%%%%%%%%%%%%%%%%%%%%%%%%%%%%%%%%%%%%%%%%%%%%%%%%%%%%%%%%%%%%%%%%%

One needs an algorithm to find all the progenitors $M_2^i$ ($i=1 \sim
N $) for $M_1$ which satisfy both the mass conservation $M_1 =
\sum_{i=1}^N M_2^i$ and equation~(\ref{eq:eps-num}). In reality, this
cannot be solved without the knowledge of the \textit{joint}
conditional probability function for all $M_2^i$ that is in fact not
known.  Therefore a variety of empirical prescriptions/tricks have
been proposed so far \citep[for instance]{KW93, SK99, SL99}. Our
algorithm basically attempts to select all the progenitors
sequentially as long as $M_2^i$ is larger than the resolution mass at
$z_2$, $M_{\rm res}(z_2)$ and the total mass satisfies
%%%%%%%%%%%%%%%%%%%%%%%%%%%%%%%%%%%%%%%%%%%%%%%%%%%%%%%%%%%%%%%%%%%
\begin{eqnarray}
  \label{eq:massconserve}
  \sum_{i=1}^N M_2^i < M_1 - \Delta M_{\rm acc}(<M_{\rm res}) .
\end{eqnarray}
%%%%%%%%%%%%%%%%%%%%%%%%%%%%%%%%%%%%%%%%%%%%%%%%%%%%%%%%%%%%%%%%%%%
The value of the minimal mass $M_{\rm res}(z_2)$ is chosen so that its
virial temperature is $10^4$K below which the gas cooling rate becomes
substantially low. This mass resolution is important so as not to
underestimate the cold gas fraction even at high redshifts.  Those
halos smaller than $M_{\rm res}(z_2)$ are not separately treated in
the merging tree, but they are collectively accounted for as an
accretion component. This is why the expression
(\ref{eq:massconserve}) has the additional term defined as
%%%%%%%%%%%%%%%%%%%%%%%%%%%%%%%%%%%%%%%%%%%%%%%%%%%%%%%%%%%%%%%%%%%
\begin{eqnarray}
\label{eq:accreted_mass}
  \Delta M_{\rm acc}(<M_{\rm res})
  = \int_{0}^{M_{\rm res}} dM_2 M_2 \frac{dN}{dM_2}(M_2,z_2|M_1,z_1) .
\end{eqnarray}
%%%%%%%%%%%%%%%%%%%%%%%%%%%%%%%%%%%%%%%%%%%%%%%%%%%%%%%%%%%%%%%%%%%

We need to trace merger trees from $z=0$ to sufficiently high
redshifts (we choose $z=30$) when cooling and (non-gravitational)
heating of gas are virtually not yet important. This is, however,
practically impossible because the number of progenitors becomes
progressively larger at higher redshifts. Therefore we decided to
prepare two different sets of merger trees at low redshifts ($z=0$ to
$z=7$) and at high redshifts ($z=7$ to $z=30$), and trace each halo of
the former at $z=7$ by using the tree of its statistical counterpart
of the latter.  The detail of this procedure is described in the
Appendix.

First we construct a merging tree realization starting at $z=0$ back
to $z=7$. Then at $z=7$, we compute masses of all the existing halos,
$M_{\mathrm{halo},i}(z=7)$,  and begin to trace their history down to
$z=0$.  For each halo at the later timestep  $z(<7)$, we search for a
halo whose mass $M_{\rm halo}(z)$ first exceeds twice the mass of its
most massive progenitor at $z=7$. If this is the case, the halo is
assigned its formation epoch as $z_{\rm f}=z$.  This procedure is
repeated for all halos at any timestep with respect to the mass of the
most massive progenitor at its formation epoch. In this way, we assign
the sequence of formation epochs for all halos along the merging tree
realization down to $z=0$.

At its formation epoch, each halo in the merger tree is assumed to
obey the density profile of dark matter:
%%%%%%%%%%%%%%%%%%%%%%%%%%%%%%%%%%%%%%%%%%%%%%%%%%%%%%%%%%%%%%%%%%%%%%
\begin{equation}
  \label{eq:nfw}
  \rho_{\rm halo}(r;M)=
  \left\{
    \begin{array}{cl}
      \displaystyle
      \frac{\overline{\rho}(z) \, \delta_{\mathrm{c}}(M)}
      {(r/r_{\mathrm{s}})(1+r/r_{\mathrm{s}})^2}
      & r< r_{\rm vir} \\ 
      \displaystyle
      0 & r>r_{\rm vir},
    \end{array}
  \right. 
\end{equation}
%%%%%%%%%%%%%%%%%%%%%%%%%%%%%%%%%%%%%%%%%%%%%%%%%%%%%%%%%%%%%%%%%%%%%%
where $\overline\rho(z) \equiv \Omega_{\mathrm{M}} \rho_\mathrm{c0}
(1+z)^3$ is the mean density of the universe at $z$,
$\rho_\mathrm{c0}$ is the present critical density,
$\delta_{\mathrm{c}}(M)$ is the characteristic density excess, and
$r_{\mathrm{s}}(M)$ indicates the scale radius of the halo
\citep{nfw96}. The virial radius $r_{\rm vir}$ is defined according to
the spherical collapse model as
%%%%%%%%%%%%%%%%%%%%%%%%%%%%%%%%%%%%%%%%%%%%%%%%%%%%%%%%%%%%%%%%%%%%%%%%%
\begin{equation}
  r_{\rm vir}(M)
  \equiv \left(\frac{3M}{4\pi\overline{\rho} \Delta_{\rm nl}}
  \right)^{1/3}, 
  \label{eq: r_vir}
\end{equation}
%%%%%%%%%%%%%%%%%%%%%%%%%%%%%%%%%%%%%%%%%%%%%%%%%%%%%%%%%%%%%%%%%%%%%%%%%
and the approximation for the critical overdensity $\Delta_{\rm nl}=
\Delta_{\rm nl}(\Omega_{\mathrm{M}}, \Omega_{\Lambda})$ can be found
in \citet{KS96}.  The two parameters $r_{\mathrm{s}}$ and $r_{\rm
  vir}$ are related in terms of the concentration parameter:
%%%%%%%%%%%%%%%%%%%%%%%%%%%%%%%%%%%%%%%%%%%%%%%%%%%%%%%%%%%%%%%%
\begin{equation}
  c(M,z) \equiv \frac{r_{\rm vir}(M,z)}{r_{\mathrm{s}}(M,z)} 
  =\frac{8.0}{1+z}\; \left(
    \frac{M}{1.4\times 10^{14}\; h_{70}^{-1}M_{\odot}}
  \right)^{-0.13} , 
  \label{eq: concentration}
\end{equation}
%%%%%%%%%%%%%%%%%%%%%%%%%%%%%%%%%%%%%%%%%%%%%%%%%%%%%%%%%%%%%%%%
where the second equality implies its empirical fitting function
\citep{Bullock01,oguri01,mshimizu03}. Finally we impose the condition
that the total mass inside $r_{\rm vir}$ is equal to $M_{\rm vir}$
which relates $\delta_{\mathrm{c}}$ to $c$ as
%%%%%%%%%%%%%%%%%%%%%%%%%%%%%%%%%%%%%%%%%%%%%%%%%%%%%%%%%%%%%%%%
\begin{equation}
  \delta_{\mathrm{c}} = 
\frac{\Delta_{\rm nl}}{3} \frac{c^3}{\ln(1+c)  -c/(1+c)}. 
\end{equation}
%%%%%%%%%%%%%%%%%%%%%%%%%%%%%%%%%%%%%%%%%%%%%%%%%%%%%%%%%%%%%%%%

We assume that the above profile for each halo does not change until
the next formation epoch for the corresponding descendant halo. At the
next formation epoch, the halo profile according to equation
(\ref{eq:nfw}) is rebuilt with the mass at that epoch.

%%%%%%%%%%%%%%%%%%%%%%%%%%%%%%%%%%%%%%%%%%%%%%%%%%%%%%%%%%%%%%%%%%%%%%
\section{Physical Models for Non-Gravitational Heating}
%%%%%%%%%%%%%%%%%%%%%%%%%%%%%%%%%%%%%%%%%%%%%%%%%%%%%%%%%%%%%%%%%%%%%%

In this paper, we focus on two physical models for non-gravitational
heating of ICM, supernova energy feedback and thermalization of jets
of radio galaxies. Their basic pictures are briefly described here.

\subsection{Energy feedback from supernovae \label{sec:SN}}

To roughly estimate the order of the available energy from supernovae,
consider the following qualitative picture. If a halo with baryonic
gas mass of $M_{\rm gas}$ cools and forms stars of mass $\Delta M_*
\equiv f_* M_{\rm gas}$, the extra heating energy is estimated as
$\eta_{\scriptscriptstyle\rm SN} f_* M_{\rm gas} \overline{E_{\rm
    K}}$, where $\eta_{\scriptscriptstyle\rm SN}$ is  the number of
supernova events per unit mass of stars and $E_\mathrm{K}$ is a
typical supernova kinetic energy.  The value of
$\eta_{\scriptscriptstyle\rm SN}$ may be computed from the adopted
stellar IMF (initial mass function).  In the case of the Salpeter IMF
$dN/d\ln M\propto M^{-x}$ with $x=1.35$ over $0.1\;M_{\odot} < M <
125\;M_{\odot}$, one obtains
%%%%%%%%%%%%%%%%%%%%%%%%%%%%%%%%%%%%%%%%%%%%%%%%%%%%%%%%%%%%%%%%%%%%%
\begin{eqnarray}
\eta_{\scriptscriptstyle\mathrm{SN}} 
= \frac{\displaystyle \int_{8M_\odot}^{125M_\odot} \frac{dN}{dM} \, dM}
{\displaystyle \int_{0.1M_\odot}^{125M_\odot} M \frac{dN}{dM} \, dM}
\approx  0.007 \;M_{\odot}^{-1} ,
\end{eqnarray}
%%%%%%%%%%%%%%%%%%%%%%%%%%%%%%%%%%%%%%%%%%%%%%%%%%%%%%%%%%%%%%%%%%%%%
if only Type II supernovae (corresponding to stars with
$M>8\;M_{\odot}$) are considered as the heating source. We adopt
$\overline{E_{\rm K}}= 10^{51}$erg.  Then the extra heating energy per
gas particle is
%%%%%%%%%%%%%%%%%%%%%%%%%%%%%%%%%%%%%%%%%%%%%%%%%%%%%%%%%%%%%%%%%%%%%
\begin{eqnarray}
\frac{\Delta E_{\scriptscriptstyle\mathrm{SN}}}{N_{\rm gas}}
&\approx& \epsilon_{\scriptscriptstyle\mathrm{SN}} 
\frac{\eta_{\scriptscriptstyle\mathrm{SN}}
\Delta M_* \overline{E_{\rm K}}}{M_{\rm gas}/m_\mathrm{p}}
= \epsilon_{\scriptscriptstyle\mathrm{SN}}
\eta_{\scriptscriptstyle\mathrm{SN}}
 f_* m_\mathrm{p} \overline{E_{\rm K}} \cr
&\approx& 3 \epsilon_{\scriptscriptstyle\mathrm{SN}} f_* 
\left(\frac{\eta_{\scriptscriptstyle\mathrm{SN}}}{0.007 \;M_{\odot}^{-1}}
\right)
\left(\frac{\overline{E_{\rm K}}}{10^{51} {\rm erg}}
\right) {\rm keV/particle} ,
\end{eqnarray}
%%%%%%%%%%%%%%%%%%%%%%%%%%%%%%%%%%%%%%%%%%%%%%%%%%%%%%%%%%%%%%%%%%%%%
where $m_\mathrm{p}$ is the proton mass. This is approximately the
amount of energy that is required to account for the
$L_{\mathrm{X}}$-$T$ relation of clusters as discussed in the previous
literature, depending on the specific values of $f_*$,
$\eta_{\scriptscriptstyle\mathrm{SN}}$, $\overline{E_{\rm K}}$, and
$\epsilon_{\scriptscriptstyle\mathrm{SN}}$.

Admittedly the estimates of these values are fairly uncertain.  Recent
studies of the IMF (\cite{Kroupa01}, \yearcite{Kroupa02}) indicate
that the $x$ of  the IMF is about 1.3 for $M > 0.5\;M_{\odot}$ and 0.3
for $0.08\;M_{\odot} < M < 0.5\;M_{\odot}$. In this case,
$\eta_{\scriptscriptstyle\mathrm{SN}} =0.013 \;M_{\odot}^{-1}$, a
factor of two larger than the above value.  Furthermore the IMF is
likely to be time-dependent, and
$\eta_{\scriptscriptstyle\mathrm{SN}}$ (and/or the star formation
efficiency $f_*$) may be signicantly larger at high redshifts given
the \textit{WMAP} implication for the early cosmic reionization.  Also
the value of $\overline{E_{\rm K}}$ may have a large scatter and the
efficiency of the energy to thermalize the ICM is not clear.  Thus our
strategy is to adopt the fiducial values of those parameters for
definiteness, and represent all the uncertainty by the overall
amplitude parameter $\epsilon_{\scriptscriptstyle\mathrm{SN}}$. Our
fiducial set of model parameters corresponds to
$\epsilon_{\scriptscriptstyle\mathrm{SN}}=1$, but we explore a broad
range $0<\epsilon_{\scriptscriptstyle\mathrm{SN}}<10$ to reflect the
uncertainty discussed above.

\subsection{Jets of radio galaxies \label{sec:RG}}

The other important heating source that we consider in this paper is
the energy input by AGNs.  Since the radiation from AGNs are very
inefficient in thermalizing the ICM, we consider a population of AGNs
which have powerful jets, Type II of the Fanaroff-Riley radio galaxies
(FR II); Type I of the FR galaxies is known to have less powerful jets
and we neglect the contribution of the latter for simplicity
\citep{Ensslin97, vs99, Ensslin00, wu00, IS01, nath02}.

To proceed further we need the luminosity function of the FR~II radio
galaxies. \citet{Willott01} proposed three different evolution models.
Specifically we adopt the intermediate evolution model (model C) which
consists of two terms for low-luminosity and high-luminosity radio
source populations:
%%%%%%%%%%%%%%%%%%%%%%%%%%%%%%%%%%%%%%%%%%%%%%%%%%%%%%%%%%%%%%%%%%%%%
\begin{equation}
  \frac{d n_{\mathrm{RG}}(L_{151}, z)}{d \log_{10}L_{151}} =
  \frac{d n_{\mathrm{RG,L}}}{d \log_{10}L_{151}}
  +\frac{d n_{\mathrm{RG,H}}}{d \log_{10}L_{151}},
  \label{eq:LFbody}
\end{equation}
%%%%%%%%%%%%%%%%%%%%%%%%%%%%%%%%%%%%%%%%%%%%%%%%%%%%%%%%%%%%%%%%%%%%%
where $L_{151}$ is luminosity of radio galaxies observed at $151$MHz.
The two terms are assumed to have the following forms:
%%%%%%%%%%%%%%%%%%%%%%%%%%%%%%%%%%%%%%%%%%%%%%%%%%%%%%%%%%%%%%%%%%%%%
\begin{equation}
  \frac{dn_{\mathrm{RG,L}}}{d\log_{10}L_{151}}=n_{\mathrm{l0}}
  \left(\frac{L_{151}}{L_{\mathrm{l\ast}}}\right)^{-\alpha_{\mathrm{l}}} 
  \exp\left(-\frac{L_{151}}{L_{\mathrm{l\ast}}}\right)\times
  \left\{
    \begin{array}{cl}
      \displaystyle
      (1+z)^{k_{\mathrm{l}}} & z < z_{\mathrm{l0}}\\
      \displaystyle
      (1+z_{\mathrm{l0}})^{k_{\mathrm{l}}} & z \geq z_{\mathrm{l0}},
    \end{array}\right.
  \label{eq:LFlowlum}
\end{equation}
%%%%%%%%%%%%%%%%%%%%%%%%%%%%%%%%%%%%%%%%%%%%%%%%%%%%%%%%%%%%%%%%%%%%%
and
%%%%%%%%%%%%%%%%%%%%%%%%%%%%%%%%%%%%%%%%%%%%%%%%%%%%%%%%%%%%%%%%%%%%%
\begin{equation}
  \frac{dn_{\mathrm{RG,H}}}{d\log_{10}L_{151}}=n_{\mathrm{h0}}
  \left(\frac{L_{151}}{L_{\mathrm{h\ast}}}\right)^{-\alpha_{\mathrm{h}}} 
  \exp\left(-\frac{L_{\mathrm{h\ast}}}{L_{151}}\right)\times
  \left\{\begin{array}{cl}
      \displaystyle
      \exp
      \left[
        -\frac{1}{2}
        \left(
          \frac{z-z_{\mathrm{h0}}}{z_{\mathrm{h1}}}
        \right)^{2}
      \right]& z < z_{\mathrm{h0}}\\
      \displaystyle
      \exp
      \left[
        -\frac{1}{2}
        \left(
          \frac{z-z_{\mathrm{h0}}}{z_{\mathrm{h2}}}
        \right)^{2}
      \right] & z \geq z_{\mathrm{h0}}.
    \end{array}\right.
  \label{eq:LFhighlum}
\end{equation}
%%%%%%%%%%%%%%%%%%%%%%%%%%%%%%%%%%%%%%%%%%%%%%%%%%%%%%%%%%%%%%%%%%%%%
The original values of the parameters in the above equations are
derived for $\Omega_{\mathrm{M}}=0$, $\Omega_{\Lambda}=0$ and
$H_{0}=50\;\mathrm{km}\; \mathrm{s}^{-1}\; \mathrm{Mpc}^{-1}$
\citep{Willott01} which are summarized in  table~\ref{tbl:RGparam}. We
convert their luminosity function  to that relevant for our
cosmological model ($\Omega_{\mathrm{M}}=0.3$, $\Omega_{\Lambda}=0.7$
and $h_{70}=1$) using equation~(14) in \citet{Willott01}.
%%%%%%%%%%%%%%%%%%%%%%%%%%%%%%%%%%%%%%%%%%%%%%%%%%%%%%%%%%%%%%%%%%%%%%
\begin{table}[thb]
  \caption{The values of the parameters for luminosity function of
    radio galaxies. \label{tbl:RGparam}}
  \begin{center}
    \begin{tabular}{ccccccccccc}\hline\hline
      $n_{\mathrm{l0}}/h_{70}^{3}\;\mathrm{Mpc}^{-3}$&
      $\alpha_{\mathrm{l}}$&
      $L_{\mathrm{l\ast}}/h_{70}^{-2} \; \mathrm{W}\;
      \mathrm{Hz}^{-1}\; \mathrm{sr}^{-1}$&
      $z_{\mathrm{l0}}$&
      $k_{\mathrm{l}}$\\
      $8.23\times 10^{-8}$&
      $0.586$&
      $1.54\times 10^{26}$&
      $0.710$&
      $3.48$&\\\hline\hline
      $n_{\mathrm{h0}}/h_{70}^{3}\;\mathrm{Mpc}^{-3}$&
      $\alpha_{\mathrm{h}}$&
      $L_{\mathrm{h\ast}}/h_{70}^{-2} \; \mathrm{W}\;
      \mathrm{Hz}^{-1}\; \mathrm{sr}^{-1}$&
      $z_{\mathrm{h0}}$&
      $z_{\mathrm{h1}}$&
      $z_{\mathrm{h2}}$\\
      $4.80\times 10^{-7}$&
      $2.42$&
      $1.25\times 10^{27}$&
      $2.03$&
      $0.568$&
      $0.956$\\\hline
    \end{tabular}
  \end{center}
\end{table}
%%%%%%%%%%%%%%%%%%%%%%%%%%%%%%%%%%%%%%%%%%%%%%%%%%%%%%%%%%%%%%%%%%%%%%

It still remains to find the average relation between the radio
luminosity $L_{151}$ and the kinetic power in jet $L_{\mathrm{j}}$ of
the population of radio galaxies.  On the basis of a physical model
combined with observational constraints of hot spot advance speeds and
spectral ages for FR II radio galaxies, \citet{Willott99} derived
%%%%%%%%%%%%%%%%%%%%%%%%%%%%%%%%%%%%%%%%%%%%%%%%%%%%%%%%%%%%%%%%%%%%%%
\begin{equation}
  \label{eq:Lorg}
  L_{\rm j} =1.5\times 10^{45} \, f_{\mathrm{j}}
  \left(
    \frac{L_{151}}{5.1\times 10^{27} \;h_{70}^{-2} \;\mathrm{W}
      \;\mathrm{Hz}^{-1} \;\mathrm{sr}^{-1}} 
  \right)^{6/7}\;h_{70}^{-2}\;\mathrm{erg}\;\mathrm{s}^{-1},
\end{equation}
%%%%%%%%%%%%%%%%%%%%%%%%%%%%%%%%%%%%%%%%%%%%%%%%%%%%%%%%%%%%%%%%%%%%%%
where $f_{\mathrm{j}}$ is a fudge factor reflecting uncertainties of
the physical condition inside the jet lobes (e.g., departures from
equipartition, proton and low-energy electron content).
\citet{Willott99} suggested $f_{\rm j} = 1 \sim 20$, and we use
$f_{\mathrm{j}}=10$ following other independent studies (e.g.,
\cite{LG01}; \cite{HW00}; \cite{BR00}; see also \S3.2 of \cite{IS01}).
Moreover \citet{IS01} showed that equation~(\ref{eq:Lorg})
underpredicts the observed jet luminosities  \citep{Rawlings92} by an
order of magnitude (see figure~1 of \cite{IS01}).

Therefore our model adopts the relation of
%%%%%%%%%%%%%%%%%%%%%%%%%%%%%%%%%%%%%%%%%%%%%%%%%%%%%%%%%%%%%%%%%%%%%%%
\begin{equation}
  \label{eq:Ljet}
  L_{\mathrm{j}}=1.5\times 10^{47}
  \left(
    \frac{L_{151}}{5.1\times 10^{27} \;h_{70}^{-2} \;\mathrm{W}
      \;\mathrm{Hz}^{-1} \;\mathrm{sr}^{-1}} 
  \right)^{6/7}\;h_{70}^{-2}\;\mathrm{erg}\;\mathrm{s}^{-1}.
\end{equation}
%%%%%%%%%%%%%%%%%%%%%%%%%%%%%%%%%%%%%%%%%%%%%%%%%%%%%%%%%%%%%%%%%%%%%%%
In order to obtain the total jet energy from $L_{\rm j}$, one needs a
typical value of the lifetime of jets, $t_{\mathrm{life}}$.  Using a
simple model of \citet{Falle91} combined with the data of 7C Redshift
Survey and 3CRR sample, \citet{Willott99} derived a constraint
$2\times 10^{6}\;\mathrm{yr} \lesssim t_{\mathrm{life}} \lesssim
10^{8}\;\mathrm{yr}$ for $L_{151}\gtrsim 5.1\times 10^{27}
\;h_{70}^{-2} \;\mathrm{W} \;\mathrm{Hz}^{-1} \;\mathrm{sr}^{-1}$.  So
we adopt $t_{\mathrm{life}} = 10^{7}\; \mathrm{yr}$ for all radio
galaxies just for simplicity and definiteness.  The above set of
parameters yields a value of the extra heating energy per gas particle
for a halo of gas mass $M_{\rm gas}$ hosting one FR II radio galaxy:
%%%%%%%%%%%%%%%%%%%%%%%%%%%%%%%%%%%%%%%%%%%%%%%%%%%%%%%%%%%%%%%%%%%%%
\begin{eqnarray}
\label{eq:E_RG}
\frac{\Delta E_{\scriptscriptstyle\mathrm{RG}}}{N_{\rm gas}}
&=& \epsilon_{\scriptscriptstyle\mathrm{RG}} 
\frac{L_{\rm j} t_{\rm life}}{M_{\rm gas}/m_p} \cr
&\approx& 2 \epsilon_{\scriptscriptstyle\mathrm{RG}}
\left(\frac{L_{\rm j}}{1.5\times10^{47}{\rm erg/s}}\right)
\left(\frac{t_{\rm life}}{10^7 {\rm year}}\right) 
\left(\frac{10^{13} M_\odot}{M_{\rm gas}}\right) 
{\rm keV/particle} .
\end{eqnarray}
%%%%%%%%%%%%%%%%%%%%%%%%%%%%%%%%%%%%%%%%%%%%%%%%%%%%%%%%%%%%%%%%%%%%%
\citet{IS01} suggest that the efficiency of the thermalization of the
jet kinetic energy may be $0.25 \sim 0.4$ if the work done by jets
stops when the pressures inside and outside of jets are equal to each
other.  Again we introduce a fudge factor,
$\epsilon_{\scriptscriptstyle\mathrm{RG}}$, to represent overall
uncertainty in our fiducial set of physical parameters.

Equation~(\ref{eq:E_RG}) indicates that the FR II radio galaxies may
be another potential candidate for the non-gravitational heating
sources of ICM.  We emphasize, however, the significant difference
between the supernova energy feedback and the jet heating; while the
former is ubiquitous and always accompanies the star formation, the
latter is a relatively rare event.   The number density of the FR II
radio galaxies, $n_{\scriptscriptstyle\mathrm{RG}}(z)$, is $\sim
2\times10^{-7}h_{70}^3$Mpc$^{-3}$ at $z=0$ [c.f.,
equation~(\ref{eq:nrgz})]. Thus one FR II per dark matter mass of
$\bar{\rho}(0)/\overline{n_{\rm\scriptscriptstyle RG}}(z=0) \sim
2\times10^{17}h_{70}^{-1}M_\odot$.  Over the cosmic age of
$10^{10}$yr, we may have $10^{10} {\rm yr}/t_{\rm life}$ generations
of radio galaxies, and thus we expect a dark halo of mass $\sim
2\times10^{14}(10^7 {\rm yr}/t_{\rm life}) h_{70}^{-1}M_\odot$ hosts
one FR II radio galaxy on average. This implies that a typical rich
cluster has been heated by the jet only for a short duration in the
past.  Thus the jet heating needs to be simulated in a stochastic
fashion as in our Monte-Carlo approach.  In contrast, the amount of
the supernova feedback is basically proportional to the halo mass, and
thus can be taken into account statistically.  In other words, the jet
heating may induce significant variations in the observed
$L_{\mathrm{X}}$-$T$ relation of galaxy clusters, and even produce a
population of outliers from the mean relation.

%%%%%%%%%%%%%%%%%%%%%%%%%%%%%%%%%%%%%%%%%%%%%%%%%%%%%%%%%%%%%%%%%%%%%%
\section{Thermal and metallicity evolution of baryonic gas inside dark
  matter halos \label{sec:gasevo}}
%%%%%%%%%%%%%%%%%%%%%%%%%%%%%%%%%%%%%%%%%%%%%%%%%%%%%%%%%%%%%%%%%%%%%%

Given the above two major scenarios for non-gravitational heating of
ICM, we describe our detailed models to trace thermal and metallicity
evolution of baryonic gas inside dark matter halos.

\subsection{Gas cooling}

We assume that the hot gas is isothermal with temperature $T_{\rm
  gas}$ and in hydrostatic equilibrium under the gravitational
potential of the dark halo.  Since we adopt equation~(\ref{eq:nfw})
for the dark halo profile, the corresponding gas density profile is
analytically expressed as follows \citep{SSM98}:
%%%%%%%%%%%%%%%%%%%%%%%%%%%%%%%%%%%%%%%%%%%%%%%%%%%%%%
\begin{eqnarray}
  \label{eq:rhogas}
  \rho_{\mathrm{hot}}(r) = \rho_{\mathrm{hot},0} \;
  \exp[-Bf(r/r_{\mathrm{s}})], 
\end{eqnarray}
%%%%%%%%%%%%%%%%%%%%%%%%%%%%%%%%%%%%%%%%%%%%%%%%%%%%%%
where
%%%%%%%%%%%%%%%%%%%%%%%%%%%%%%%%%%%%%%%%%%%%%%%%%%%%%%
\begin{eqnarray}
  B &=& \frac{2c}{\ln(1+c)-c/(1+c)}
  \frac{T_{\mathrm{vir}}}{T_{\mathrm{gas}}},\\
  f(x) &=& 1-\frac{1}{x}\ln (1+x) .
\end{eqnarray}
%%%%%%%%%%%%%%%%%%%%%%%%%%%%%%%%%%%%%%%%%%%%%%%%%%%%%%
We define the virial temperature $T_{\mathrm{vir}}$ as
%%%%%%%%%%%%%%%%%%%%%%%%%%%%%%%%%%%%%%%%%%%%%%%%%%%%%%%%%%%%%%%%%
\begin{equation}
  k_{\mathrm{B}}T_{\mathrm{vir}}=\frac{1}{2}\mu m_{\mathrm{p}}
  \frac{GM_{\mathrm{vir}}}{r_{\mathrm{vir}}},
  \label{eq:virtemp}
\end{equation}
%%%%%%%%%%%%%%%%%%%%%%%%%%%%%%%%%%%%%%%%%%%%%%%%%%%%%%%%%%%%%%%%%
where $G$ is the gravitational constant, $\mu m_{\mathrm{p}}$ is a
mean molecular weight ($\mu \simeq 0.6$), $k_{\mathrm{B}}$ is the
Boltzmann constant.  The amplitude $\rho_{\rm hot, 0}$ is computed so
as to reproduce the total hot gas mass in each halo when integrating
equation~(\ref{eq:rhogas}) within $r=r_{\rm vir}$.

%%%% gas cooling

Once the gas profile is specified, we compute the cooling radius
$r_{\rm cool}$ within which the gas cools at each redshift $z$ or the
cosmic time $t(z)$. The gas cooling time-scale at a radius $r$ is
given by
%%%%%%%%%%%%%%%%%%%%%%%%%%%%%%%%%%%%%%%%%%%%%%%%%%%%%%%%%%%%%%%%%%%%%%
\begin{equation}
  \label{eq:tcool}
  t_{\mathrm{cool}}(r)=\frac{3}{2}
  \frac{\rho_{\mathrm{hot}}(r)}{\mu m_{\mathrm{p}}}
  \frac{k_{\mathrm{B}}T_{\mathrm{gas}}}
  {\Lambda(T_{\mathrm{gas}},Z)[n_{\mathrm{H}}(r)]^{2}}, 
\end{equation}
%%%%%%%%%%%%%%%%%%%%%%%%%%%%%%%%%%%%%%%%%%%%%%%%%%%%%%%%%%%%%%%%%%%%%%
where $\Lambda(T_{\mathrm{gas}},Z)$ is the cooling rate of gas with
temperature $T_{\mathrm{\mathrm{gas}}}$ and metallicity $Z$, and
$n_{\mathrm{H}}(r)$ is the number density of hydrogen (including both
neutral and ionized). We compute the evolution of $Z$ for each halo
simultaneously, and apply the relevant cooling rate
$\Lambda(T_{\mathrm{gas}},Z)$ by using tables in
\citet{sutherlanddopita}. For physically reasonable profiles,
equation~(\ref{eq:tcool}) is a monotonically increasing function of
$r$, and  $r_{\rm cool}$ is computed from the condition:
%%%%%%%%%%%%%%%%%%%%%%%%%%%%%%%%%%%%%%%%%%%%%%%%%%%%%%%%%%%%%%%%%%%%%%
\begin{equation}
  t_{\rm cool}(r_{\rm cool})= \tau_{\rm cool}(z)
\equiv  t(z) - t(z_{\rm f}(M_{\rm f})) .
\end{equation}
%%%%%%%%%%%%%%%%%%%%%%%%%%%%%%%%%%%%%%%%%%%%%%%%%%%%%%%%%%%%%%%%%%%%%%

While $\tau_{\rm cool}(z)$ in the above equation \textit{conceptually}
denotes the elapse time that gas in each halo can spend for cooling,
its exact definition is fairly uncertain.  In the second equality, we
adopted the definition of \citet{mshimizu02} based on the following
simple picture; when a halo of mass $M_{\rm f}$ \textit{forms} at the
formation redshift $z_{\rm f}$, its hot gas is supposed to reach the
profile [equation~(\ref{eq:rhogas})] instantaneously.  This defines
the origin of the cooling time for the halo, and $\tau_{\rm cool}$ is
set to the elapsed cosmic time since $z_{\rm f}$.  In the subsequent
timesteps, we neglect the change in the hot gas profile, gas
temperature, and metallicity, even if the halo mass $M$ grows due to
mergers.  At each formation epoch of the host halo, its hot gas
profile is reset to the profile [equation~(\ref{eq:rhogas})]
corresponding to the new mass and temperature, and the origin of
$\tau_{\rm cool}$ is replaced by that at the new formation epoch.

We apply this procedure for all halos, and the cold gas in each
progenitor halo is simply accumulated according to the merger trees.
For halos that do not have any direct progenitor at higher redshifts,
we assume that they consist of hot gas only with no excess energy and
no metals, and set their gas temperature as the virial temperature of
the halos. The accreted gas [to represent the gas components below our
mass resolution, see equation~(\ref{eq:accreted_mass})] is always
assumed to join the hot gas component of the halo without excess
energy or metals.

\subsection{Gas heating \label{subsec:heating}}

%%%% supernova feedback

At the formation epoch of each halo (i.e., each mass doubling time to
be more specific), we compute the increase of the cold gas mass
$\Delta M_{\mathrm{cold}}$ of the halo  since the previous formation
epochs of all progenitor halos.

The increased amount of the cold gas mass $\Delta M_{\rm cold}$ is
assumed to instantaneously form stars of mass $\Delta M_* = (1-f_{\rm
  rh})\Delta M_{\rm cold}$ while the rest of it $f_{\rm rh}\Delta
M_{\rm cold}$ is reheated by the supernova energy and returns to the
hot gas of the halo.  The returned hot gas carries the metals produced
by massive stars and pollutes the hot gas associated with the halo.
Observationally, metal (especially, Fe) abundances of hot gas around
NGC~1399, a cD galaxy in the Fornax Cluster are estimated as
$1.1^{+1.3}_{-0.5}\;Z_{\odot}$ within 360 kpc \citep{Ikebe92}, and
$1.5\mbox{--}2.0\;Z_{\odot}$ within 20 kpc \citep{Buote02}.
Theoretical models of metal ejection from elliptical galaxies by
galactic wind also suggests a higher than the solar abundance (e.g.,
\cite{David90}), independently of the mass of stars at their final
stage. Therefore we assume that the metallicity of the returned hot
gas, $Z_{\mathrm{eject}}$, is equal to $2\;Z_\odot$.

%%%% radio galaxy

In addition to the supernova heating, we incorporate the energy input
by radio galaxies as follows.  First, we compute the number density,
$n_{\scriptscriptstyle\mathrm{RG}}(z)$, of the radio galaxies at $z$
using equations~(\ref{eq:LFbody}), (\ref{eq:LFlowlum}),
(\ref{eq:LFhighlum}), and table~\ref{tbl:RGparam}:
%%%%%%%%%%%%%%%%%%%%%%%%%%%%%%%%%%%%%%%%%%%%%%%%%%%%%%%%%%%%%%%%%%%%%%%%
\begin{equation}
\label{eq:nrgz}
  n_{\mathrm{RG}}(z)=\int_{\log_{10}L_{\mathrm{151,min}}} ^{\infty}
  \frac{dn_{\mathrm{RG}}}{d\log_{10}L_{151}}(L_{151}, z)\;
  d\log_{10}L_{151}.
\end{equation}
%%%%%%%%%%%%%%%%%%%%%%%%%%%%%%%%%%%%%%%%%%%%%%%%%%%%%%%%%%%%%%%%%%%%%%%%
We assume that the minimum luminosity of the radio galaxies
$L_{\mathrm{151,min}}$ is $1.6\times 10^{25}\;h_{70}^{-2} \;
\mathrm{W}\; \mathrm{Hz}^{-1}\; \mathrm{sr}^{-1}$ at
$151\;\mathrm{MHz}$, which corresponds to the FR~I/FR~II dichotomy in
the radio morphology \citep{Bicknell95,Laing96,Fabian01}.

Next, we compute the expected number,
$\overline{N_{\scriptscriptstyle\rm RG}}(M_{\rm halo},z)$, of the
radio galaxies for a halo $M_{\mathrm{halo}}$ at $z$ during each time
step of the merging tree realization $\Delta t_{\rm tree}(z)$:
%%%%%%%%%%%%%%%%%%%%%%%%%%%%%%%%%%%%%%%%%%%%%%%%%%%%%%%%%%%%%%%%%%%%%%%%
\begin{equation}
\label{eq:RGprob}
 \overline{N_{\scriptscriptstyle\rm RG}}(M_{\mathrm{halo}},z) =
  \frac{M_{\mathrm{halo}}n_{\scriptscriptstyle\mathrm{RG}}(z)} 
  {\int_{M_{\mathrm{min}}}^{\infty}
    Mn_{\scriptscriptstyle\mathrm{PS}}(M,z)dM} 
  \frac{\Delta t_{\mathrm{tree}}(z)}{t_{\mathrm{life}}} ,
\end{equation}
%%%%%%%%%%%%%%%%%%%%%%%%%%%%%%%%%%%%%%%%%%%%%%%%%%%%%%%%%%%%%%%%%%%%%%%%
where the minimum mass of halos that host radio galaxies is set as
$M_{\mathrm{min}} = 10^{12}\;M_{\odot}$.  In the above equation,
$n_{\scriptscriptstyle\mathrm{PS}}(M,z)$ is the mass function of dark
halos for which we use the Press-Schechter formula for definiteness.
Note that the last factor in equation~(\ref{eq:RGprob}) accounts for
the number of \textit{generations} of the radio galaxies during
$\Delta t_{\rm tree}(z)$.

Finally, we randomly assign the radio galaxies to each halo in merger
trees for $z<7$ (we ignore the jet heating for $z>7$) according to the
expected number. As in the above treatment of the gas, we sum up the
number of radio galaxies at each timestep but input the corresponding
jet heating only at the formation epoch of the host halo. This
procedure gives rise to the stochasticity in the heating.  The radio
luminosity of each radio galaxy is randomly assigned according to the
observed luminosity function.  We denote the amount of total energy
input at the formation epoch by $\Delta
E_{\scriptscriptstyle\mathrm{RG}}$.

\subsection{Determining the gas temperature with the non-gravitational
  heating}

%%%% temperature

We compute the total amount of excess energy $\Delta E_{\mathrm{gas}}$
for each halo at its formation epoch:
%%%%%%%%%%%%%%%%%%%%%%%%%%%%%%%%%%%%%%%%%%%%%%%%%%%%%%%%%%%%%%%%%%%%%%%%
\begin{equation}
  \label{eq:degas}
  \Delta E_{\mathrm{gas}} \equiv  E_{\mathrm{ex,prog}} + \Delta
  E_{\scriptscriptstyle\mathrm{SN}} 
  + \Delta E_{\scriptscriptstyle\mathrm{RG}} - \Delta
  E_{\mathrm{cool}}, 
\end{equation}
%%%%%%%%%%%%%%%%%%%%%%%%%%%%%%%%%%%%%%%%%%%%%%%%%%%%%%%%%%%%%%%%%%%%%%%%
where $E_{\mathrm{ex,prog}}$ is the total excess energy already stored
in all the progenitor halos, $\Delta
E_{\scriptscriptstyle\mathrm{SN}}$ and $\Delta
E_{\scriptscriptstyle\mathrm{RG}}$ indicate the excess energies from
supernova feedback and jets of radio galaxies since the last formation
epoch of the progenitor halos, and $\Delta E_{\mathrm{cool}}$ is the
loss of the excess energy due to radiative cooling.

The total energy loss $\Delta E_{\mathrm{cool}}$ is calculated as
$\Delta E_{\mathrm{cool}} =\sum_{i} e_{\mathrm{ex},i} \Delta
M_{\mathrm{cold},i}$, where $e_{\mathrm{ex},i} \equiv 3
k_{\mathrm{B}}(T_{\mathrm{gas},i}-T_{\mathrm{vir},i})/(2\mu
m_{\mathrm{p}})$  and $\Delta M_{\mathrm{cold},i}$ is the increase of
cold gas mass since the previous formation epoch of each progenitor
halo $i$.

The excess energy thermalizes the surrounding ICM as well as exerts
work against gravity (e.g., \cite{wu00}). If we neglect the
contribution of self-gravity of the gas, we can define the total
gravitational energy of the gas as follows:
%%%%%%%%%%%%%%%%%%%%%%%%%%%%%%%%%%%%%%%%%%%%%%%%%%%%%%%%%%%%%%%%%%%%%%%%
\begin{equation}
  E_{\mathrm{grav}}(T_{\mathrm{gas}})=
  \int \rho_{\mathrm{hot}}(r)\Phi_{\mathrm{halo}}(r)dV,
\end{equation}
%%%%%%%%%%%%%%%%%%%%%%%%%%%%%%%%%%%%%%%%%%%%%%%%%%%%%%%%%%%%%%%%%%%%%%%%
where $\Phi_{\mathrm{halo}}(r)$ is the gravitational potential energy
of the halo and given as
%%%%%%%%%%%%%%%%%%%%%%%%%%%%%%%%%%%%%%%%%%%%%%%%%%%%%%%%%%%%%%%%%%%%%%%%
\begin{equation}
  \Phi_{\mathrm{halo}}(r)=-4\pi Gr_{\mathrm{s}}^{2}\delta_{\mathrm{c}}
  \overline{\rho}(z)
  \left[
    \frac{\ln(1+r/r_{\mathrm{s}})}{r/r_{\mathrm{s}}}-\frac{1}{1+c}
  \right]
\end{equation}
%%%%%%%%%%%%%%%%%%%%%%%%%%%%%%%%%%%%%%%%%%%%%%%%%%%%%%%%%%%%%%%%%%%%%%%%
for the density profile of equation~(\ref{eq:nfw}).  Then the total
energy of gas of temperature $T_{\rm gas}$ is written as the sum of
the above gravitational energy and the thermal energy
$E_{\mathrm{thermal}}(T_{\mathrm{gas}})$:
%%%%%%%%%%%%%%%%%%%%%%%%%%%%%%%%%%%%%%%%%%%%%%%%%%%%%%%%%%%%%%%%%%%%%%%%
\begin{equation}
  E_{\mathrm{gas}}(T_{\mathrm{gas}})
  =E_{\mathrm{grav}}(T_{\mathrm{gas}})
  +E_{\mathrm{thermal}}(T_{\mathrm{gas}}).
\end{equation}
%%%%%%%%%%%%%%%%%%%%%%%%%%%%%%%%%%%%%%%%%%%%%%%%%%%%%%%%%%%%%%%%%%%%%%%%
Since we assume that the hot gas in each halo is isothermal,
$E_{\mathrm{thermal}}(T_{\mathrm{gas}})$ is simply equal to
$3M_{\mathrm{hot}}k_{\mathrm{B}}T_{\mathrm{gas}}/(2\mu
m_{\mathrm{p}})$, where $M_{\mathrm{hot}}$ is the mass of hot gas
component in the halo.

Now we are in a position to determine the gas temperature with the
given energy budget. In reality this is not easy as in the empirical
approach that we attempt here, since we do not know the efficiency of
even \textit{gravitational} shock heating. Therefore we first adopt a
conventional assumption in modeling galaxy clusters that the
gravitational shock heating is sufficiently effective to keep the gas
temperature $T_{\rm gas}$ equal to its virial temperature $T_{\rm
  vir}$.  Then we adopt that an additional assumption that the excess
energy $\Delta E_{\rm gas}$ is exclusively consumed to increase the
temperature beyond $T_{\rm vir}$. Thus we solve the following equation
for $T_{\rm gas}$ given $E_{\mathrm{gas}}(T_{\mathrm{vir}})$ from
gravitational shock heating and the excess energy due to
non-gravitational heating [equation~(\ref{eq:degas})]:
%%%%%%%%%%%%%%%%%%%%%%%%%%%%%%%%%%%%%%%%%%%%%%%%%%%%%%%%%%%%%%%%%%%%%%%%
\begin{equation}
  E_{\mathrm{gas}}(T_{\mathrm{gas}}) =
  E_{\mathrm{gas}}(T_{\mathrm{vir}}) 
  + \Delta E_{\mathrm{gas}}.
  \label{eq:tgas}
\end{equation}
%%%%%%%%%%%%%%%%%%%%%%%%%%%%%%%%%%%%%%%%%%%%%%%%%%%%%%%%%%%%%%%%%%%%%%%%

\subsection{Diffuse gas and the metallicity evolution}

%%%% diffuse gas

With the presence of non-gravitational heating, it occasionally
happens that the gas temperature becomes much higher than the virial
temperature of the halo, and hydrostatic equilibrium is unrealistic.
In order to take account of such cases properly, we assume that hot
gas is completely ejected from the host halo if the ratio
$T_{\mathrm{gas}}/T_{\mathrm{vir}}$ exceeds the critical ratio
$f_{\mathrm{diffuse}}$.  The ejected gas is not bound to the halo, but
assumed to exist as a \textit{diffuse} gas component with no
subsequent thermal and metallicity evolution.  The diffuse component
joins the bound hot gas component again when the descendent of the
progenitor halo increases its halo mass so that its virial temperature
$T'_{\mathrm{vir}}$ exceeds $T_{\mathrm{gas}}/f_{\mathrm{diffuse}}$ of
the diffuse gas. Then the diffuse gas returns to the hot gas carrying
the total gas energy and the metals acquired at the time when it is
ejected.  While we use the value of $f_{\mathrm{diffuse}}=10$, the
results at $z=0$ are insensitive to the choice for $10\lesssim
f_{\mathrm{diffuse}}\lesssim 100$ since the hydrostatic configuration
itself [equation~(\ref{eq:rhogas})] is very shallow and effectively
approximates a profile close to the diffuse component.

%%%% metallicity

Finally the mass of metals for each halo is traced according to
%%%%%%%%%%%%%%%%%%%%%%%%%%%%%%%%%%%%%%%%%%%%%%%%%%%%%%%%%%%%%%%%%%%%%%%%
\begin{equation}
  M_{\mathrm{Z}} \equiv M_{\mathrm{Z,prog}} + Z_{\mathrm{eject}}
  f_{\mathrm{rh}} \Delta M_{\mathrm{cold}}
  - \sum_{i}Z_{i}\Delta M_{\mathrm{cold},i},
\end{equation}
%%%%%%%%%%%%%%%%%%%%%%%%%%%%%%%%%%%%%%%%%%%%%%%%%%%%%%%%%%%%%%%%%%%%%%%%
where $M_{\mathrm{Z,prog}}$ is the total mass of metals in all the
progenitors, and $Z_{i}$ denotes the metallicity in each progenitor
halo $i$. As explained in \S\ref{subsec:heating}, we set $Z_{\rm
  eject}=2Z_\odot$.

%%%% table of parameters

In passing, we summarize in table~\ref{tbl:thermalparam} the
parameters used in modeling thermal and metallicity evolution of gas.
Our model is characterized by the three free parameters
$f_{\mathrm{rh}}$, $\epsilon_{\scriptscriptstyle\mathrm{SN}}$, and
$\epsilon_{\scriptscriptstyle\mathrm{RG}}$, which represent the
fraction of the reheated gas due to the supernova feedback, and the
(dimensionless) strengths of the supernova heating and the jet heating
that we explore in the subsequent sections.
%%%%%%%%%%%%%%%%%%%%%%%%%%%%%%%%%%%%%%%%%%%%%%%%%%%%%%%%%%%%%%%%%%%%%%
\begin{table}[thb]
  \caption{Parameters in the present model.   \label{tbl:thermalparam}}
  \begin{center}
    \begin{tabular}{ccl}\hline\hline
      symbol & adopted value & physical meaning \\ \hline
      $\tau_{\rm cool}$ & halo mass doubling time & time for cooling 
      gas in each host halo \\
      $\eta_{{\scriptscriptstyle\mathrm{SN}}}$ & $0.007\; M_{\odot}^{-1}$ &
      the number of supernovae per unit mass of stars formed out of 
      cold gas \\
      $E_\mathrm{kin}$ & $10^{51}\;\mathrm{erg}$ & kinetic energy of
      supernova explosion\\ 
      $Z_\mathrm{eject}$ & $2\;Z_\odot$ & metallicity of the ejected gas\\
      $L_{\mathrm{j}}$ & equation~(\ref{eq:Ljet}) & kinetic luminosity of 
      jets of an FR II radio galaxy\\
      $t_{\mathrm{life}}$ & $10^{7}\;\mathrm{yr}$ & life time of 
      jets of an FR II radio galaxy\\
      $f_{\mathrm{diffuse}}$ & 10 & a ratio to define the  
      diffuse gas component\\
      \hline
      $f_{\mathrm{rh}}$ & (free) & reheated gas fraction of cold gas to ICM\\
      $1-f_{\mathrm{rh}}$ & --- & star formation efficiency\\
      $\epsilon_{\scriptscriptstyle\mathrm{SN}}$ & (free) & efficiency
      of energy input by supernova feedback \\
      $\epsilon_{\scriptscriptstyle\mathrm{RG}}$ & (free) & efficiency
      of energy input by jets of radio galaxies \\
      \hline
    \end{tabular}
  \end{center}
\end{table}
%%%%%%%%%%%%%%%%%%%%%%%%%%%%%%%%%%%%%%%%%%%%%%%%%%%%%%%%%%%%%%%%%%%%%%

%%%%%%%%%%%%%%%%%%%%%%%%%%%%%%%%%%%%%%%%%%%%%%%%%%%%%%%%%%%%%%%%%%%%%%
\section{Constraints from the metallicity--temperature relation
\label{sec:frhfix}}
%%%%%%%%%%%%%%%%%%%%%%%%%%%%%%%%%%%%%%%%%%%%%%%%%%%%%%%%%%%%%%%%%%%%%%

In order to avoid working in the three-dimensional parameter space, we
attempt to put constraints on $f_{\mathrm{rh}}$ from the
metallicity--temperature relation of clusters. Since the metallicity
is produced by the reheated gas from supernova explosion, we can
neglect the effect of the radio galaxy heating
($\epsilon_{\scriptscriptstyle\mathrm{RG}}=0$) except when
$\epsilon_{\scriptscriptstyle\mathrm{SN}}$ is very small ($<0.5$) as
discussed below.  We first generate many realizations of clusters at
$z=0$ over a wide range of mass systematically changing the values of
$f_{\mathrm{rh}}$ and $\epsilon_{\scriptscriptstyle\mathrm{SN}}$. For
each cluster, we compute the gas temperature $T_{\rm gas}$ and the
metallicity $Z$ according to the procedure described in the last
section.  Also we compute its bolometric luminosity as:
%%%%%%%%%%%%%%%%%%%%%%%%%%%%%%%%%%%%%%%%%%%%%%%%%%%%%%%%%%%%%%%%%%%%%%%%%%%
\begin{equation}
  \label{eq:lx}
  L_{\mathrm{bol}}=4\pi\int^{r_{\rm vir}}_{0}\Lambda
  (T_{\rm gas}, Z) [n_{\mathrm{H}}(r)]^{2}
  r^{2}dr.
\end{equation}
%%%%%%%%%%%%%%%%%%%%%%%%%%%%%%%%%%%%%%%%%%%%%%%%%%%%%%%%%%%%%%%%%%%%%%%%%%%

From those simulated clusters, we select samples whose gas temperature
$T_{\rm gas}$ is higher than 2.5 keV \textit{and} whose bolometric
luminosity is larger than $0.1 \overline{L_{\rm bol}}(T_{\rm gas})$ at
its gas temperature, where $\overline{L_{\rm bol}}(T_{\rm gas})$ is
the mean bolometric luminosity of clusters in the observational sample
compiled by \citet{Ikebe02}. The lower limit on the luminosity is
introduced so as to incorporate the observational flux limit
approximately, but our results are insensitive the value.  For that
selected sample of simulated clusters ($N_{\rm sel}$ in total), we
compute the statistics:
%%%%%%%%%%%%%%%%%%%%%%%%%%%%%%%%%%%%%%%%%%%%%%%%%%%%%%%%%%%%%%%%%%%%%%%%%%%
\begin{eqnarray}
  \chi^{2}_{\mathrm{red}} &=&
  \frac{1}{N_{\mathrm{sel}}}
  \sum_{i=1}^{N_{\mathrm{sel}}}
  \frac{[Z_{\mathrm{model},i}-\overline{Z_{\mathrm{obs}}}
    (T_{\mathrm{model},i})]^{2}}{\sigma_{\rm obs, Z}^{2}},\\  
  \overline{Z_{\mathrm{obs}}}(T) &=& [
    0.298-0.005 (T/{\rm keV})] Z_{\odot},
\end{eqnarray}
%%%%%%%%%%%%%%%%%%%%%%%%%%%%%%%%%%%%%%%%%%%%%%%%%%%%%%%%%%%%%%%%%%%%%%%%%%%
where $Z_{\mathrm{model},i}$ and $T_{\mathrm{model},i}$ are the
metallicity and the gas temperature of $i$-th cluster in the sample
($i=1 \sim N_{\rm sel}$). The mean metallicity-temperature relation
$\overline{Z_{\mathrm{obs}}}(T)$ is fitted from the observed data of
Fe abundances of \citet{fukazawa98}. The corresponding standard
deviation in the observed data is computed as
%%%%%%%%%%%%%%%%%%%%%%%%%%%%%%%%%%%%%%%%%%%%%%%%%%%%%%%%%%%%%%%%%%%%%%%
\begin{equation}
  \sigma_{\rm obs, Z}^{2}=
  \frac{1}{N_{\mathrm{obs}}} 
  \sum_{i=1}^{N_{\mathrm{obs}}}
  [Z_{\mathrm{obs},i}-\overline{Z_{\mathrm{obs}}}(T_{\mathrm{obs},i})]^{2},
\end{equation}
%%%%%%%%%%%%%%%%%%%%%%%%%%%%%%%%%%%%%%%%%%%%%%%%%%%%%%%%%%%%%%%%%%%%%%%
where $N_{\mathrm{obs}}(=40)$ is the total number of the sample of
clusters in \citet{fukazawa98}, and $Z_{\mathrm{obs},i}$ and
$T_{\mathrm{obs},i}$ are the metallicity and the gas temperature of
the $i$-th cluster.  We obtain $\sigma_{\rm obs, Z} \approx 0.063\;
Z_{\odot}$.

%%%%%%%%%%%%%%%%%%%%%%%%%%%%%%%%%%%%%%%%%%%%%%%%%%%%%%%%%%%%%%%%%%%%%%
\begin{figure}[tbh]
  \centering \FigureFile(80mm,80mm){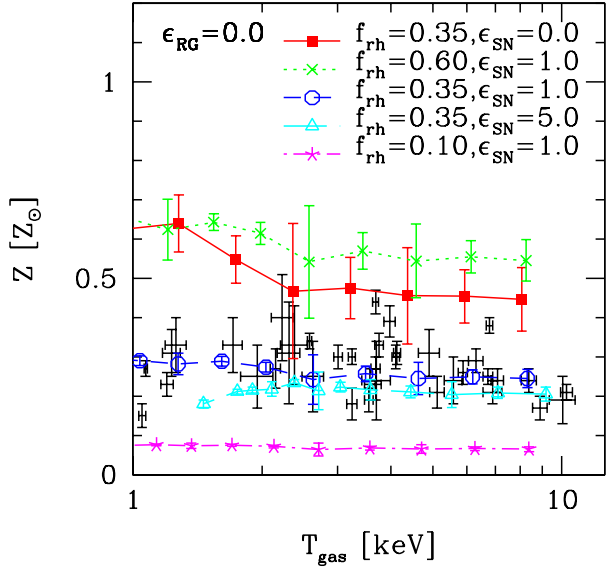}
\caption{Metallicity-temperature relations of clusters at $z=0$.
  The dots with vertical and horizontal error-bars indicate the
  observational data \citet{fukazawa98}.  The other symbols with
  vertical error-bars represent the mean and the corresponding
  standard deviation from 10 realization halos sorted according to
  their gas temperature.
\label{fig:TZexample}}
\end{figure}
%%%%%%%%%%%%%%%%%%%%%%%%%%%%%%%%%%%%%%%%%%%%%%%%%%%%%%%%%%%%%%%%%%%%%%

\begin{figure}[tbh]
  \centering \FigureFile(80mm,80mm){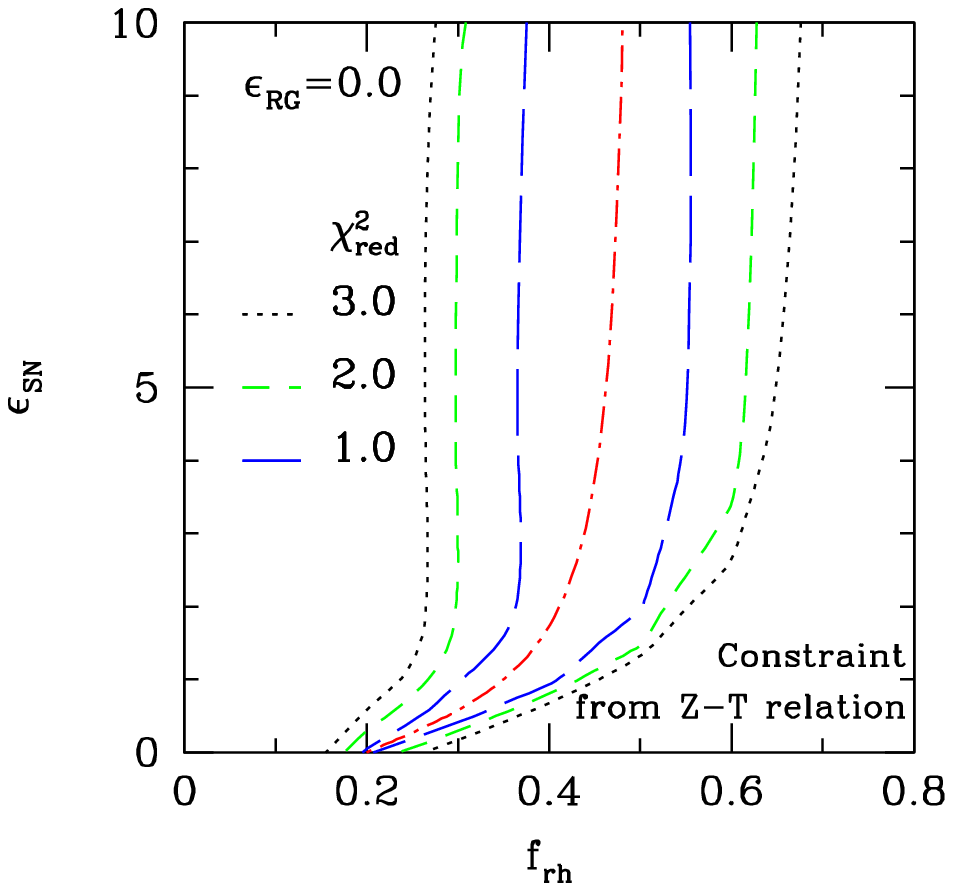}
  \caption{Constraints on ($f_{\mathrm{rh}}$,
    $\epsilon_{\scriptscriptstyle\mathrm{SN}}$) from the observed
    metallicity-temperature relation by \citet{fukazawa98}.  The
    contours of $\chi^{2}_{\mathrm{red}}=1$, 2 and 3 are plotted.  The
    best-fit empirical relation [equation~(\ref{eq:esn})] that
    reproduces the observational metallicity-temperature is plotted in
    dot-dashed line.
    \label{fig:ContMetTemp}}
\end{figure}

Examples of the predicted $Z$--$T_{\rm gas}$ relation are shown in
figure~\ref{fig:TZexample}.  As expected, the metallicity is mainly
determined by the reheated gas fraction $f_{\rm rh}$.  The contours of
$\chi^{2}_{\mathrm{red}}$ on the $f_{\mathrm{rh}}$ --
$\epsilon_{\scriptscriptstyle\mathrm{SN}}$ plane (for
$\epsilon_{\scriptscriptstyle\mathrm{RG}}=0$) are plotted in
figure~\ref{fig:ContMetTemp}.  Note that
$\epsilon_{\scriptscriptstyle\mathrm{SN}}$ represents the
\textit{efficiency} of the supernova energy feedback in our model, and
the supernova rate $\eta_{\scriptscriptstyle\rm SN}$ is assumed to be
independent of the value of
$\epsilon_{\scriptscriptstyle\mathrm{SN}}$. This is why our model has
metallicity evolution even when
$\epsilon_{\scriptscriptstyle\mathrm{SN}}=0$ as long as the hot gas
cools and forms cold gas and stars.  The resulting metallicity
increases as $f_{\rm rh}$, but decreases as
$\epsilon_{\scriptscriptstyle\mathrm{SN}}$ since the star formation
rate is suppressed due to the stronger energy input from the
supernovae.  When $\epsilon_{\scriptscriptstyle\mathrm{SN}}\gtrsim 3$,
however, the result becomes almost insensitive to
$\epsilon_{\scriptscriptstyle\mathrm{SN}}$ and simply determined by
the value of $f_{\rm rh}$ alone.  In this regime, heating is so strong
and the suppression of  gas cooling is almost saturated. Thus the
further increase of the energy feedback does not change the result.

%%%%%%%%%%%%%%%%%%%%%%%%%%%%%%%%%%%%%%%%%%%%%%%%%%%%%%%%%%%%%%%%%%%%%%
%%%%%%%%%%%%%%%%%%%%%%%%%%%%%%%%%%%%%%%%%%%%%%%%%%%%%%%%%%%%%%%%%%%%%%

Just for the convenience of the analyses below, we obtained an
empirical fit to the $f_{\mathrm{rh}}$ --
$\epsilon_{\scriptscriptstyle\mathrm{SN}}$ relation from
figure~\ref{fig:ContMetTemp} which reproduces the observational
metallicity-temperature relation:
%%%%%%%%%%%%%%%%%%%%%%%%%%%%%%%%%%%%%%%%%%%%%%%%%%%%%%%%%%%%%%%%%%%%%%
\begin{equation}
  \label{eq:esn}
  \begin{array}{cc}
    \displaystyle
    \epsilon_{\scriptscriptstyle\mathrm{SN}} \sim \tan 
    \left[
      \frac{5\pi}{3}(f_{\mathrm{rh}}-0.2) 
    \right] & (f_{\mathrm{rh}}\lesssim 0.5).
  \end{array}
\end{equation}
%%%%%%%%%%%%%%%%%%%%%%%%%%%%%%%%%%%%%%%%%%%%%%%%%%%%%%%%%%%%%%%%%%%%%%
It is encouraging that the resulting range, $0.2 \lesssim
f_{\mathrm{rh}} \lesssim 0.5$, is roughly consistent with other models
of evolution of elliptical galaxies (\cite{David90},
\yearcite{David91a}, \yearcite{David91b}; \cite{Elbaz95}).

We also repeat the above analysis for
$\epsilon_{\scriptscriptstyle\mathrm{RG}} \not= 0$, and do not find
significant difference as long as
$\epsilon_{\scriptscriptstyle\mathrm{SN}}\gtrsim 0.6$ because  heating
by radio galaxies is rather stochastic as noted in \S\ref{sec:RG} and
does not affect the mean properties of clusters compared with the
supernova feedback (see also figure~\ref{fig:tz} below).

%%%%%%%%%%%%%%%%%%%%%%%%%%%%%%%%%%%%%%%%%%%%%%%%%%%%%%%%%%%%%%%%%%%%%%
\section{Constraints from the luminosity--temperature relation}
\label{sec:epsfix}
%%%%%%%%%%%%%%%%%%%%%%%%%%%%%%%%%%%%%%%%%%%%%%%%%%%%%%%%%%%%%%%%%%%%%%

Let us move next to deriving constraints on the
$\epsilon_{\scriptscriptstyle\mathrm{SN}}$ --
$\epsilon_{\scriptscriptstyle\mathrm{RG}}$ plane from the observed
$L_{\mathrm{bol}}$ -- $T_{\rm gas}$ relation of clusters.  In doing
so, we fix the value of $f_{\rm rh}$ as a function of
$\epsilon_{\scriptscriptstyle\mathrm{SN}}$ so as to reproduce the
metallicity -- temperature relation. To be specific, we invert
equation~(\ref{eq:esn}) and adopt
%%%%%%%%%%%%%%%%%%%%%%%%%%%%%%%%%%%%%%%%%%%%%%%%%%%%%%%%%%%%%%%%%%
\begin{equation}
  \label{eq:frh}
  f_{\mathrm{rh}} = \frac{3}{5\pi}\arctan
  (\epsilon_{\scriptscriptstyle\mathrm{SN}}) +0.2 .
\end{equation}
%%%%%%%%%%%%%%%%%%%%%%%%%%%%%%%%%%%%%%%%%%%%%%%%%%%%%%%%%%%%%%%%%%

We would like to confront our simulated cluster samples at $z=0$ with
the observational data combined by \citet{Ikebe02}.  The latter is a
flux-limited sample of the flux limit $S_{\mathrm{lim}}=2\times
10^{-11} \;\mathrm{erg}\;\mathrm{s}^{-1}\;\mathrm{cm}^{-2}$ in the
0.1--2.4 keV band with spectroscopically measured temperature for
$T_{\mathrm{gas}}>2.5\;\mathrm{keV}$.  The number of these clusters is
$\tilde{N}_{\mathrm{obs}}=52$ in total. Since our simulated clusters
are not assigned the distance, we cannot construct the corresponding
flux-limited sample in reality. Thus we select those simulated
clusters whose gas temperature $T_{\rm gas}$ is higher than 2.5 keV
\textit{and} whose bolometric luminosity is larger than $0.1
\overline{L_{\rm bol}}(T_{\rm gas})$ as in the previous section.

%%%%%%%%%%%%%%%%%%%%%%%%%%%%%%%%%%%%%%%%%%%%%%%%%%%%%%%%%%%%%%%%%%%%%%
\begin{figure}[tbh]
  \centering \FigureFile(80mm,80mm){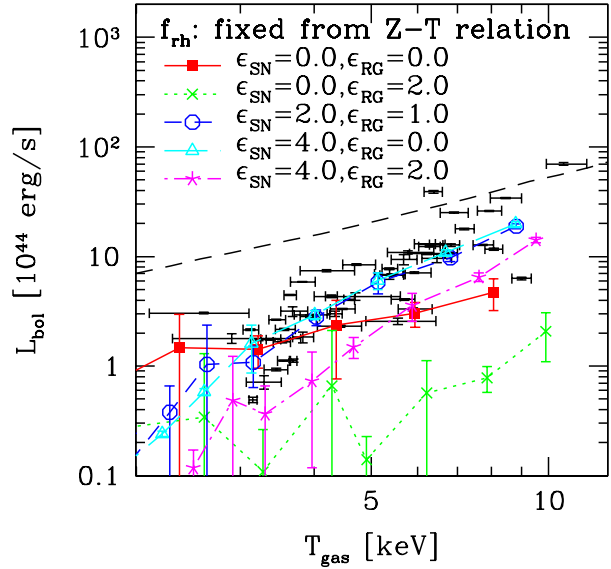}
\caption{Luminosity-temperature relations of clusters at $z=0$.  The
  dots with vertical and horizontal error-bars indicate observational
  data from \citet{Ikebe02}.  The other symbols with vertical
  error-bars represent the mean and the corresponding standard
  deviation from 10 realization halos sorted according to their gas
  temperature.  The dashed line indicates the `self-similar' relation,
  in which all baryons in the halo have
  $T_{\mathrm{gas}}=T_\mathrm{vir}$.  \label{fig:LTexample}}
\end{figure}
%%%%%%%%%%%%%%%%%%%%%%%%%%%%%%%%%%%%%%%%%%%%%%%%%%%%%%%%%%%%%%%%%%%%%%

For such selected samples of simulated clusters, we compute
%%%%%%%%%%%%%%%%%%%%%%%%%%%%%%%%%%%%%%%%%%%%%%%%%%%%%%%%%%%%%%%%%%
\begin{eqnarray}
  \tilde{\chi}^{2}_{\mathrm{red}} &=&
  \frac{1}{N_{\mathrm{sel}}}
  \sum_{i=1}^{N_{\mathrm{sel}}}
  \frac{[\log_{10}(L_{\mathrm{model},i}/10^{44}\; \mathrm{erg}\;
    \mathrm{s}^{-1}) - 
    \overline{F_{\rm obs, log L}}(T_{\mathrm{model},i})]^{2}} 
  {\sigma_\mathrm{log L}^{2}},\\
  \overline{F_{\rm obs, log L}}(T) &=& -1.17+2.81\log_{10}
    (T/{\rm keV}) ,
\end{eqnarray}
%%%%%%%%%%%%%%%%%%%%%%%%%%%%%%%%%%%%%%%%%%%%%%%%%%%%%%%%%%%%%%%%%%
where $L_{\mathrm{model},i}$ and $T_{\mathrm{model},i}$ are the
bolometric luminosity and the gas temperature of the $i$-th simulated
cluster ($i=1 \sim N_{\rm sel} $).  The function $\overline{F_{\rm
    obs, log L}}(T)$ is our best-fit `$\log_{10} L_{\mathrm{bol}}$ --
$\log_{10} T_{\rm gas}$' relation to the data of \citet{Ikebe02}.  The
corresponding standard deviation in the observed data is computed as
%%%%%%%%%%%%%%%%%%%%%%%%%%%%%%%%%%%%%%%%%%%%%%%%%%%%%%%%%%%%%%%%%%
\begin{equation}
  \sigma_{\mathrm{log L}}^{2}=
  \frac{1}{\tilde{N}_{\mathrm{obs}}}
  \sum_{i=1}^{\tilde{N}_{\mathrm{obs}}}
  \left[\log_{10}\left(\frac{L_{\mathrm{obs},i}}{10^{44}\;
        \mathrm{erg}\; \mathrm{s}^{-1}}\right) -
    \overline{F_{\rm obs, log L}}(T_{\mathrm{obs},i})\right]^{2}, 
\end{equation}
%%%%%%%%%%%%%%%%%%%%%%%%%%%%%%%%%%%%%%%%%%%%%%%%%%%%%%%%%%%%%%%%%%
where $L_{\mathrm{obs},i}$ and $T_{\mathrm{obs},i}$ are the bolometric
luminosity and the gas temperature of the $i$-th observed cluster
($i=1 \sim \tilde{N}_{\rm obs}$). We obtain $\sigma_{\mathrm{log
    L}}=0.24$.

In order to perform the fit over a wide dynamic range of the
bolometric luminosities, we decided to do so in logarithmic scales
instead of in linear scales. Thus the variable
$\tilde{\chi}^{2}_{\mathrm{red}}$ does not obey the standard
$\chi^2$-distribution.  We use the value to examine simply the degree
of the goodness-of-fit between our model and the observation, and do
not intend to assign any statistical significance in a strict sense.
%%%%%%%%%%%%%%%%%%%%%%%%%%%%%%%%%%%%%%%%%%%%%%%%%%%%%%%%%%%%%%%%%%%%%%
\begin{figure}[tbh]
  \centering \FigureFile(80mm,80mm){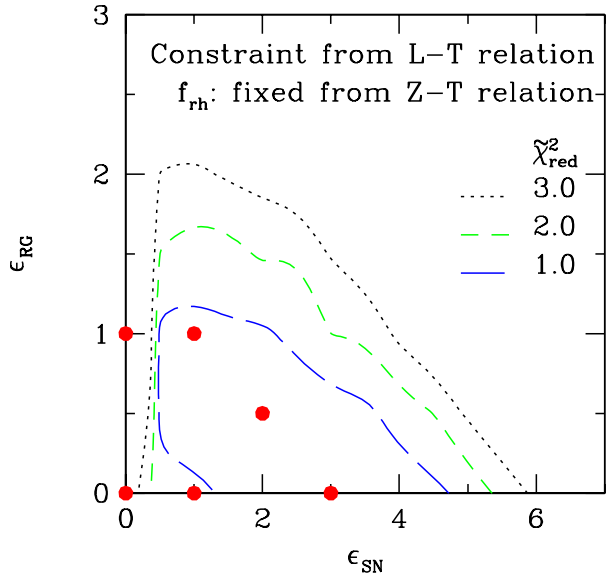}
\caption{Constraints on ($\epsilon_{\scriptscriptstyle\mathrm{SN}}$,
  $\epsilon_{\scriptscriptstyle\mathrm{RG}}$) from the observed
  luminosity-temperature relation by \citet{Ikebe02}. The contours of
  $\tilde{\chi}^{2}_{\mathrm{red}}=1$, 2, and 3 are plotted.  We fix
  the value of $f_{\mathrm{rh}}$ according to equation~(\ref{eq:frh}).
  The six filled circles indicate the parameter sets whose statistical
  properties are examined in detail later (\S\ref{sec:results}).
    \label{fig:ContLumTemp}}
\end{figure}
%%%%%%%%%%%%%%%%%%%%%%%%%%%%%%%%%%%%%%%%%%%%%%%%%%%%%%%%%%%%%%%%%%%%%%

Figure~\ref{fig:LTexample} shows examples of our model
$L_{\mathrm{X}}$--$T$ relation against the observational data.  This
clearly indicates that the $L_{\mathrm{X}}$--$T$ relation is very
sensitive to the efficiency of the non-gravitational heating processes
as expected.  The case of purely gravitational heating
($\epsilon_{\scriptscriptstyle\mathrm{SN}}=
\epsilon_{\scriptscriptstyle\mathrm{RG}}=0$; solid line with solid
squares) reproduces the slope of the analytical self-similar
prediction (dashed line).  Our model incorporating gas cooling,
however, results in a significant fraction of cold gas in all clusters
(see figure~\ref{fig:fgas} below), and the amplitude is lower than the
latter in this case. The change of the $L_{\mathrm{X}}$--$T$ relation
is induced through those of the cold gas fraction and the hot gas
density profile due to non-gravitational heating.

Figure~\ref{fig:ContLumTemp} plots the contours of
$\tilde{\chi}^{2}_{\mathrm{red}}$ on
$\epsilon_{\scriptscriptstyle\mathrm{SN}}$ --
$\epsilon_{\scriptscriptstyle\mathrm{RG}}$ plane. As is well known,
the supernova feedback (or more strictly, any uniformly heating source
of the ICM) is needed to account for the observed
$L_{\mathrm{X}}$--$T$ relation.  The jet of radio galaxies alone
cannot provide the required degree of heating for clusters. Keeping
the constraint in mind, we will examine various statistical properties
of clusters in the next section.  In particular we focus on the six
sets of parameters, $(\epsilon_{\scriptscriptstyle\mathrm{SN}},
\epsilon_{\scriptscriptstyle\mathrm{RG}}) = (0.0, 0.0)$, $(0.0, 1.0)$,
$(1.0, 0.0)$, $(1.0, 1.0)$, $(3.0, 0.0)$, and $(2.0, 0.5)$, plotted in
figure~\ref{fig:ContLumTemp}.  Incidentally \citet{kay03} conducted
similar studies using cosmological hydrodynamic simulations.  While we
compute the SN gas temperature/entropy from the amount of the heating
energy [equation (\ref{eq:tgas})], they explicitly assume its
temperature. Thus direct comparison with their simulations needs
caution, but still our reading of their table 1 is that the
significant amount of SN heating (i.e., their $\epsilon >1$) is
required to simultaneously account for the $L_{\mathrm{X}}$--$T$
relation and the hot gas fraction of ICM. This is consistent with our
above successful parameter set
$(\epsilon_{\scriptscriptstyle\mathrm{SN}},
\epsilon_{\scriptscriptstyle\mathrm{RG}}) = (3.0, 0.0)$ as we show
below.

%%%%%%%%%%%%%%%%%%%%%%%%%%%%%%%%%%%%%%%%%%%%%%%%%%%%%%%%%%%%%%%%%%%%%%
\section{Statistical properties of simulated clusters \label{sec:results}}
%%%%%%%%%%%%%%%%%%%%%%%%%%%%%%%%%%%%%%%%%%%%%%%%%%%%%%%%%%%%%%%%%%%%%%

%%%%%%%%%%%%%%%%%%%%%%%%%%%%%%%%%%%%%%%%%%%%%%%%%%%%%%%%%%%%%%%%%%%%%%
\begin{figure}[tbh]
  \centering \FigureFile(80mm,80mm){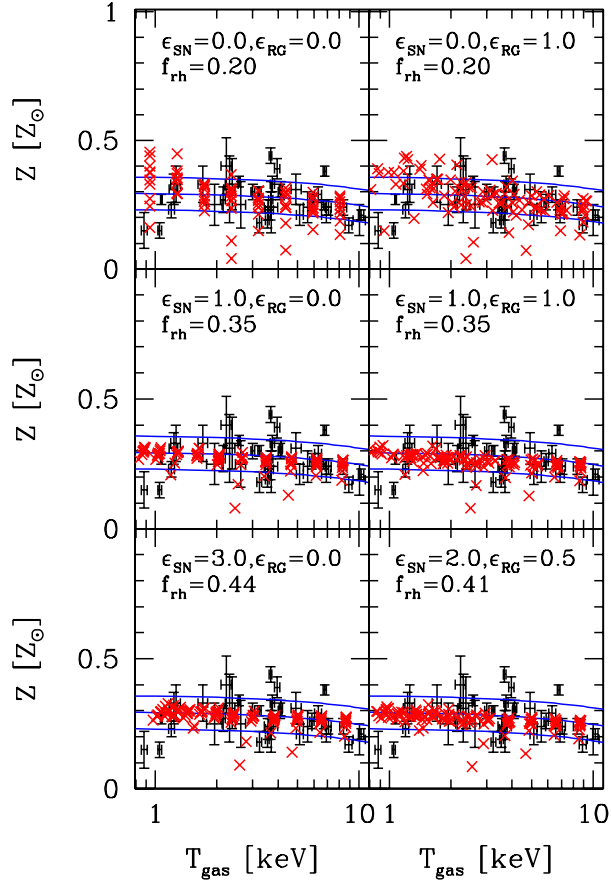}
\caption{Scatter plots of the metallicity-temperature relation of
  clusters at $z=0$.  Our model predictions are plotted in crosses,
  and the observed Fe abundances \citep{fukazawa98} are plotted in
  dots with error-bars.  The solid lines indicate the mean
  observational metallicity-temperature relation and the corresponding
  $\pm\;  1\sigma$ dispersions.  \label{fig:tz}}
\end{figure}
%%%%%%%%%%%%%%%%%%%%%%%%%%%%%%%%%%%%%%%%%%%%%%%%%%%%%%%%%%%%%%%%%%%%%%

%\subsection{Metallicity-Temperature Relation \label{sec:tz}}

The metallicity-temperature relations of our simulated clusters in the
six models are plotted in figure~\ref{fig:tz}. We adopt
equation~(\ref{eq:frh}) for $f_{\rm rh}$ even in the case of
$\epsilon_{\scriptscriptstyle\mathrm{RG}}\ne 0$ (right panels).  Since
the values of $f_{\rm rh}$ are chosen so as to reproduce the $Z$ --
$T_{\rm gas}$ relation for $\epsilon_{\scriptscriptstyle\mathrm{RG}}=
0$, the agreement in left panels is just by construction. Right panels
in figure~\ref{fig:tz} confirm that the non-vanishing
$\epsilon_{\scriptscriptstyle\mathrm{RG}}$ does not affect the mean
relation as mentioned in \S\ref{sec:frhfix}, but mainly add scatters
around the mean due to the stochastic nature of heating by radio
galaxies in our model. Note, however, that it is probably premature to
compare the dispersions around the mean even if tempting. It is not
clear to what extent the observational data reflect the intrinsic
scatter rather than merely the observational errors. Furthermore our
modeling is very simplified and ignores various possible sources for
the dispersions; we assumed constant
$\eta_{\scriptscriptstyle\mathrm{SN}}$, $f_{\rm rh}$,
$\epsilon_{\scriptscriptstyle\mathrm{SN}}$,
$\epsilon_{\scriptscriptstyle\mathrm{RG}}$, and so on which are likely
to be dependent on the environment. In this sense, we expect that our
model predictions systematically underestimate the real scatters.
This remark should apply also to the other statistical properties
discussed in this section.

%%%%%%%%%%%%%%%%%%%%%%%%%%%%%%%%%%%%%%%%%%%%%%%%%%%%%%%%%%%%%%%%%%%%%%
\begin{figure}[tbh]
  \centering \FigureFile(80mm,80mm){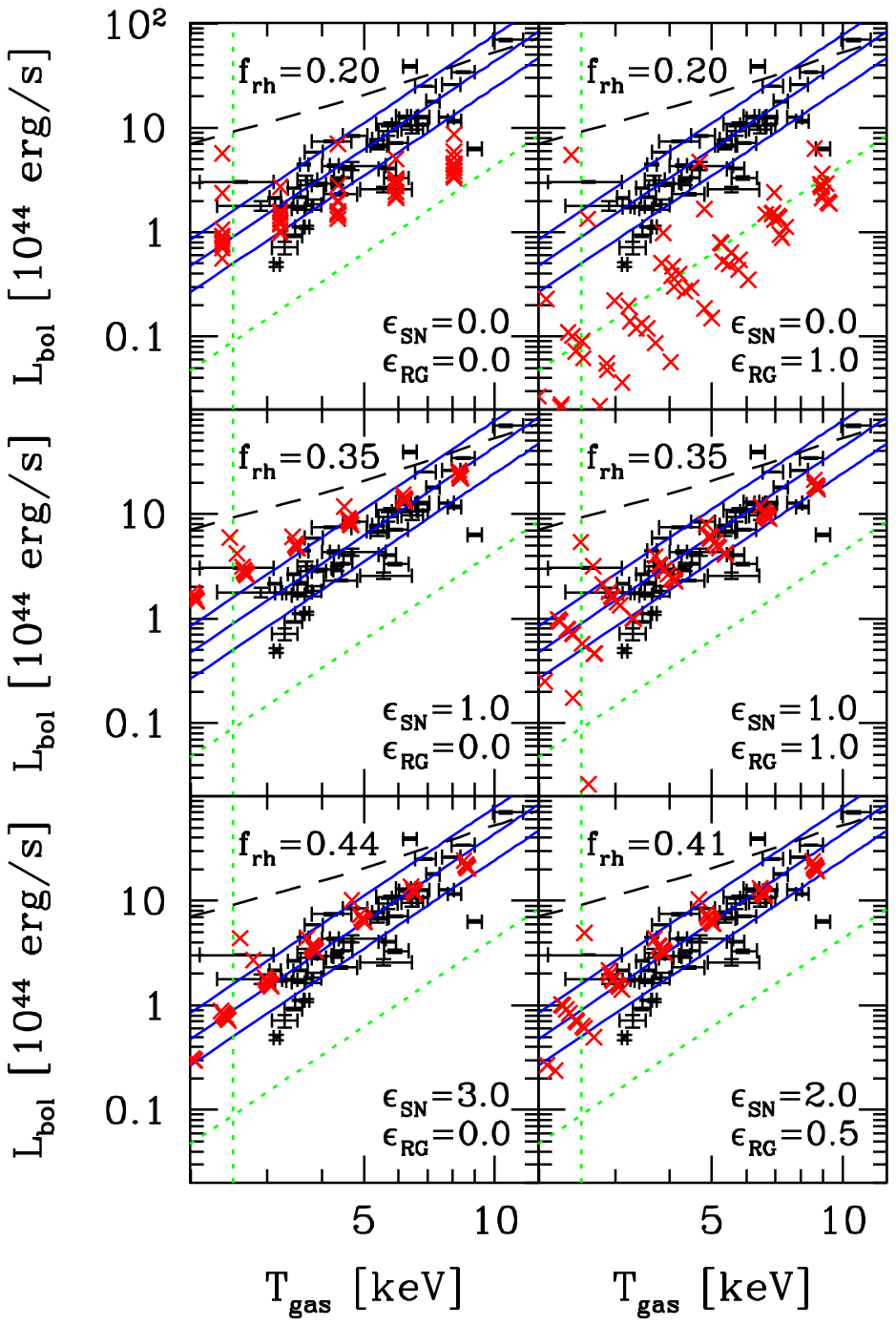}
\caption{Scatter plots of the luminosity-temperature relation of
  clusters at $z=0$.  Our model predictions are plotted in crosses,
  and the observational data \citep{Ikebe02} are plotted in dots with
  error-bars.  The solid lines indicate the mean observational
  luminosity-temperature relation and the corresponding $\pm\;
  1\sigma$ dispersions.  The dotted lines represent our selection
  criteria; $T_{\mathrm{gas}}>2.5\;\mathrm{keV}$ and 0.1 times the
  mean $L_{\rm bol}(T_{\rm gas})$.  The dashed line indicates the
  `self-similar' relation corresponding to figure~\ref{fig:LTexample}.
  \label{fig:lt}}
\end{figure}
%%%%%%%%%%%%%%%%%%%%%%%%%%%%%%%%%%%%%%%%%%%%%%%%%%%%%%%%%%%%%%%%%%%%%%

Figure~\ref{fig:lt} shows the degree of the goodness-of-fit of the
luminosity--temperature relation in detail for the six sets of model
parameters. Top panels indicate that the supernova feedback is indeed
essential in explaining the $L_{\mathrm{bol}}$-$T$ relation.  Middle
and bottom panels suggest that the jet heating is relatively
insignificant if efficiency of the supernova feedback is high.
Although the $\tilde{\chi}^{2}_{\mathrm{red}}$-statistics in
figure~\ref{fig:ContLumTemp} may indicate a marginally acceptable fit
in the case of $(\epsilon_{\scriptscriptstyle\mathrm{SN}},
\epsilon_{\scriptscriptstyle\mathrm{RG}}) = (1.0, 0.0)$, further
heating (either by the jet or by much higher star formation efficiency
as empirically proposed in \S\ref{sec:highsn} below) is indeed
preferred.  It is interesting to note that the jet heating sometimes
produces an \textit{X-ray dark cluster}.  This corresponds to a
cluster in the middle-right panel of figure~\ref{fig:lt}; the mass and
the temperature of the cluster is $7.9\times 10^{13}M_{\odot}$, and
2.7 keV while it is very X-ray faint ($L_{\mathrm{bol}} \sim 7\times
10^{42}\; \mathrm{erg}\; \mathrm{s}^{-1}$). This is an illustrative
example which reflects the stochastic nature of the jet heating.

\begin{figure}[tbh]
  \centering \FigureFile(80mm,80mm){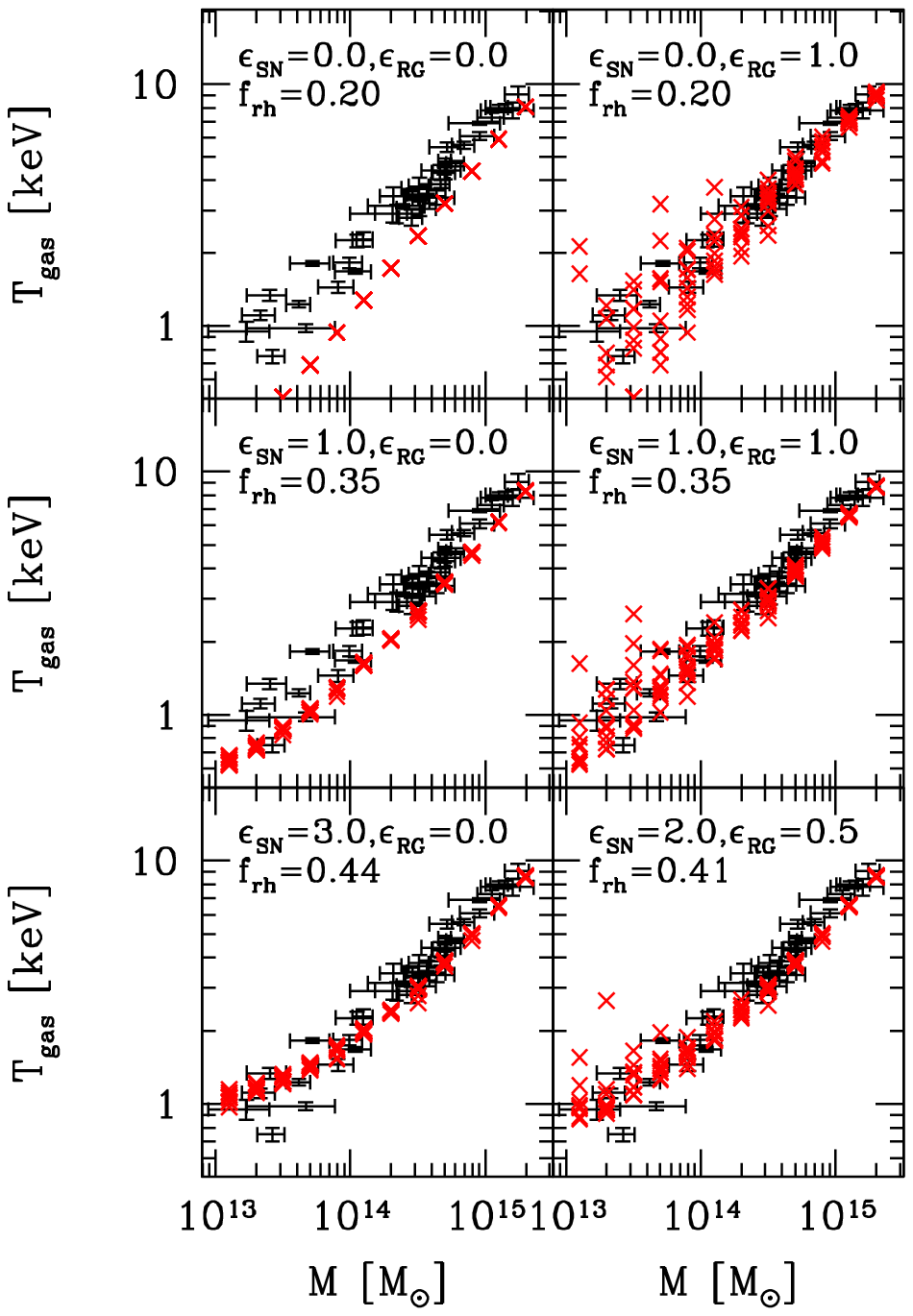}
\caption{Scatter plots of the mass-temperature relation of clusters at
  $z=0$. Our model predictions are plotted in crosses. The dots with
  vertical and horizontal error-bars indicate the observational data
  points of \citet{finogu} after appropriate correction for the virial
  mass (c.f., Appendix A in \cite{mshimizu03}).  \label{fig:mt}}
\end{figure}

Figure~\ref{fig:mt} shows the results for the mass -- temperature
relation of the clusters, which are in good agreement with the
conclusions in figure~\ref{fig:lt}; we need strong heating sources to
match the observed $M$ - $T$ relation \citep{finogu}.  While this may
be due to the fact that the $M$ - $T$ relation is not entirely
independent of the $L_{\mathrm{bol}}$-$T$ relation (since our current
model adopts the specific density and temperature profiles as a
function of mass alone), it will provide additional insight at least
(see also \cite{mshimizu03}).

%%%%%%%%%%%%%%%%%%%%%%%%%%%%%%%%%%%%%%%%%%%%%%%%%%%%%%%%%%%%%%%%%%%%%%
%%%%%%%%%%%%%%%%%%%%%%%%%%%%%%%%%%%%%%%%%%%%%%%%%%%%%%%%%%%%%%%%%%%%%%

%%%%%%%%%%%%%%%%%%%%%%%%%%%%%%%%%%%%%%%%%%%%%%%%%%%%%%%%%%%%%%%%%%%%%%
\begin{figure}[tbh]
  \centering \FigureFile(80mm,80mm){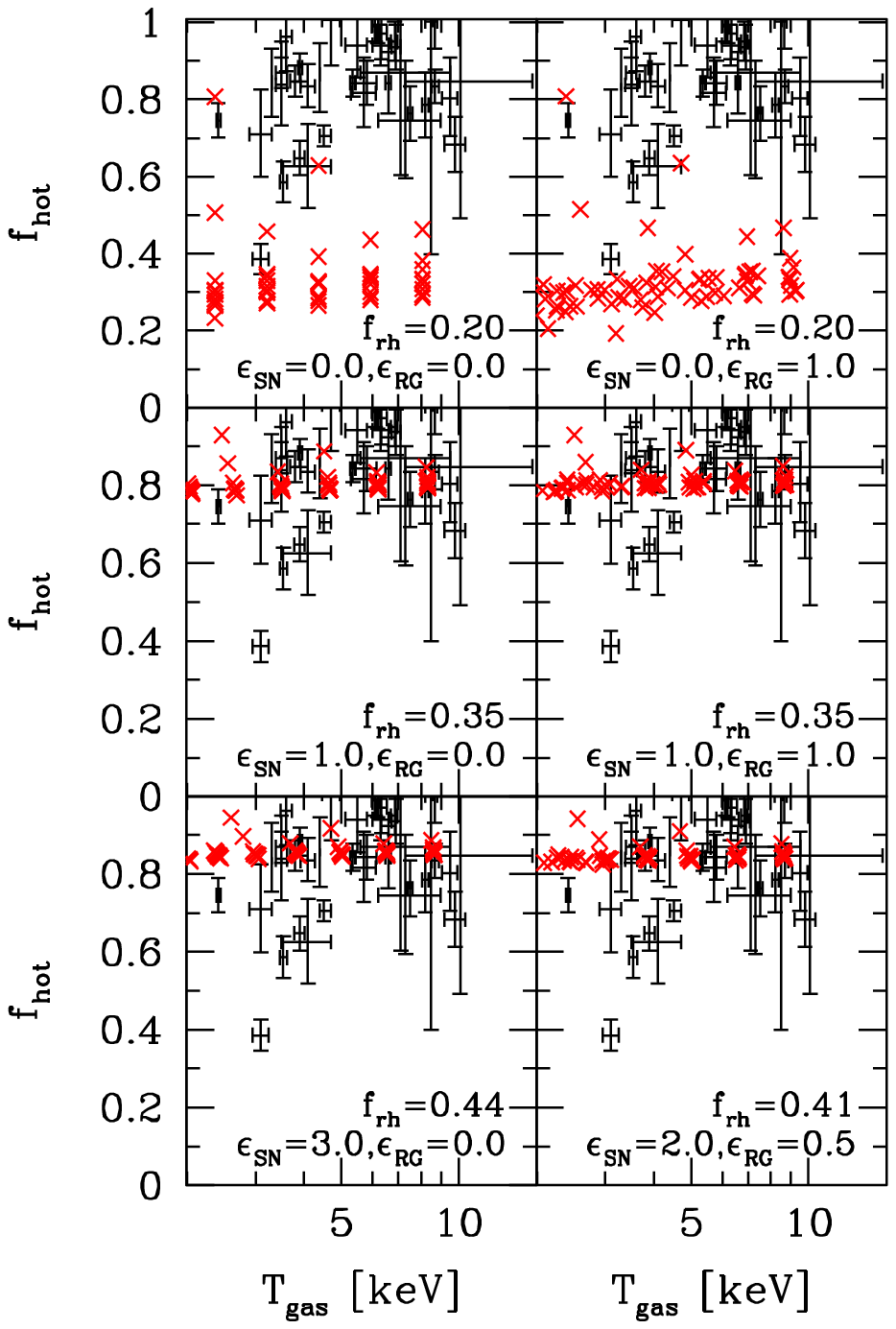}
\caption{Scatter plots of the hot gas fraction of clusters $z=0$.
  Our model predictions are plotted in crosses. The observational data
  points are taken from \citet{Mohr99}, but are multiplied by
  $\Omega_\mathrm{M}/\Omega_{\mathrm{B}}$.
\label{fig:fgas}}
\end{figure}
%%%%%%%%%%%%%%%%%%%%%%%%%%%%%%%%%%%%%%%%%%%%%%%%%%%%%%%%%%%%%%%%%%%%%%

The successful heating models also reproduce well the observed hot gas
fraction of clusters. In figure~\ref{fig:fgas}, we plot the hot gas
mass fraction in halos at $z=0$ against their gas temperature,
$f_{\mathrm{hot}}(T_{\rm gas}) \equiv M_{\rm hot}/(\Omega_\mathrm{B}
M_{\rm halo}(T_{\rm gas})  /\Omega_{\mathrm{M}})$. Again the average
value of $f_{\mathrm{hot}}(T_{\rm gas})$ is mainly controlled by the
amount of supernova energy feedback, and is fairly insensitive to the
value of $\epsilon_{\scriptscriptstyle\mathrm{RG}}$.  Note that
cooling of ICM is so efficient and without non-gravitational heating
most of the ICM remains cold.  This is another reason why a
significant amount of non-gravitational heating is required in galaxy
clusters (again not entirely independent of the luminosity --
temperature relation, though).

\citet{ponman99} find that the `core entropy', $S_{\rm c}$, of low
temperature clusters ($T_{\rm gas} \sim 1$keV) is much larger than the
self-similar model prediction, where $S_{\rm c} =
k_{\mathrm{B}}T_{\mathrm{gas}}
/[n_{\mathrm{e}}(r_{\mathrm{c}})]^{2/3}$ is defined at the core radius
$r_{\rm c}$.  This departure from the self-similarity of ICM also
supports the importance of non-gravitational heating/cooling processes
in the evolution of galaxy clusters; when some fraction of gas in the
ICM cools, the hot gas density responsible for the entropy is reduced
and the gas entropy increases.  Of course the gas entropy is increased
by the temperature rise and the resulting spreading of the core region
due to non-gravitational heating. In reality, both processes are
intimately coupled with each other and one has to trace the thermal
evolution of ICM taking account of heating and cooling in a
self-consistent fashion.  Although this seems a straightforward
project for hydrodynamic numerical simulations, it is not easy to
reliably trace the thermal evolution over the required dynamic range
between $\lesssim 10^7M_\odot$ (that can cool at $z\sim 30$
corresponding to the virial temperature of $\sim 10^4$K) and $\gtrsim
10^{16}M_\odot$ (that encloses the region surrounding a rich cluster)
given the limited mass resolution of current simulations.  This is
exactly why our current approach plays a complementary role in
studying the thermal evolution of ICM.

\begin{figure}[tbh]
  \centering \FigureFile(80mm,80mm){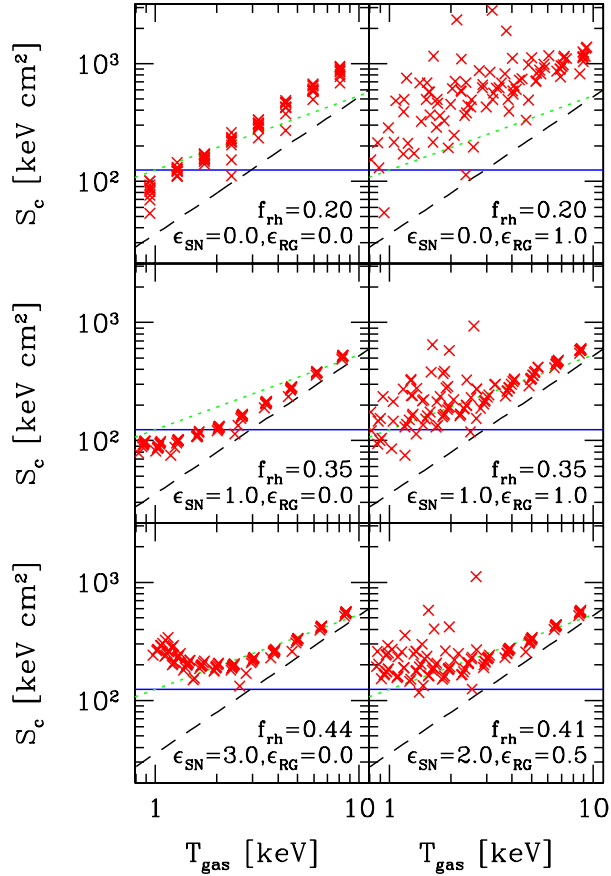}
\caption {Scatter plots of the core `entropy' of clusters at $z=0$.
  Our model predictions are plotted in crosses. The solid line
  indicates the entropy floor of $124\; h_{70}^{-1/3} \;
  \mathrm{keV}\; \mathrm{cm}^{2}$ from \citet{lloyd00}. The dotted
  line is the observational best-fit relation by \citet{ponman03}.
  The dashed line represents the self-similar model prediction.
  \label{fig:Entropy}}
\end{figure}

Figure~\ref{fig:Entropy} shows the core `entropy' against the gas
temperature of halos at $z=0$.  In the plot, we adopt a simple
definition for the core radius, $r_{\mathrm{c}}=0.1r_{\mathrm{vir}}$,
which is approximately consistent with most observed clusters.  We
estimate $n_{\mathrm{e}}(r)$ from equation~(\ref{eq:rhogas}) assuming
that the hot gas is in collisional ionization equilibrium.  Despite
such crude definitions adopted tentatively, our heating models fairly
reproduce the observational best fit relation by \citet{ponman03} and
the value of the entropy `floor' by \citet{lloyd00}.

%%%%%%%%%%%%%%%%%%%%%%%%%%%%%%%%%%%%%%%%%%%%%%%%%%%%%%%%%%%%%%%%%%%%%%
%%%%%%%%%%%%%%%%%%%%%%%%%%%%%%%%%%%%%%%%%%%%%%%%%%%%%%%%%%%%%%%%%%%%%%

\begin{figure}[tbh]
  \centering \FigureFile(80mm,80mm){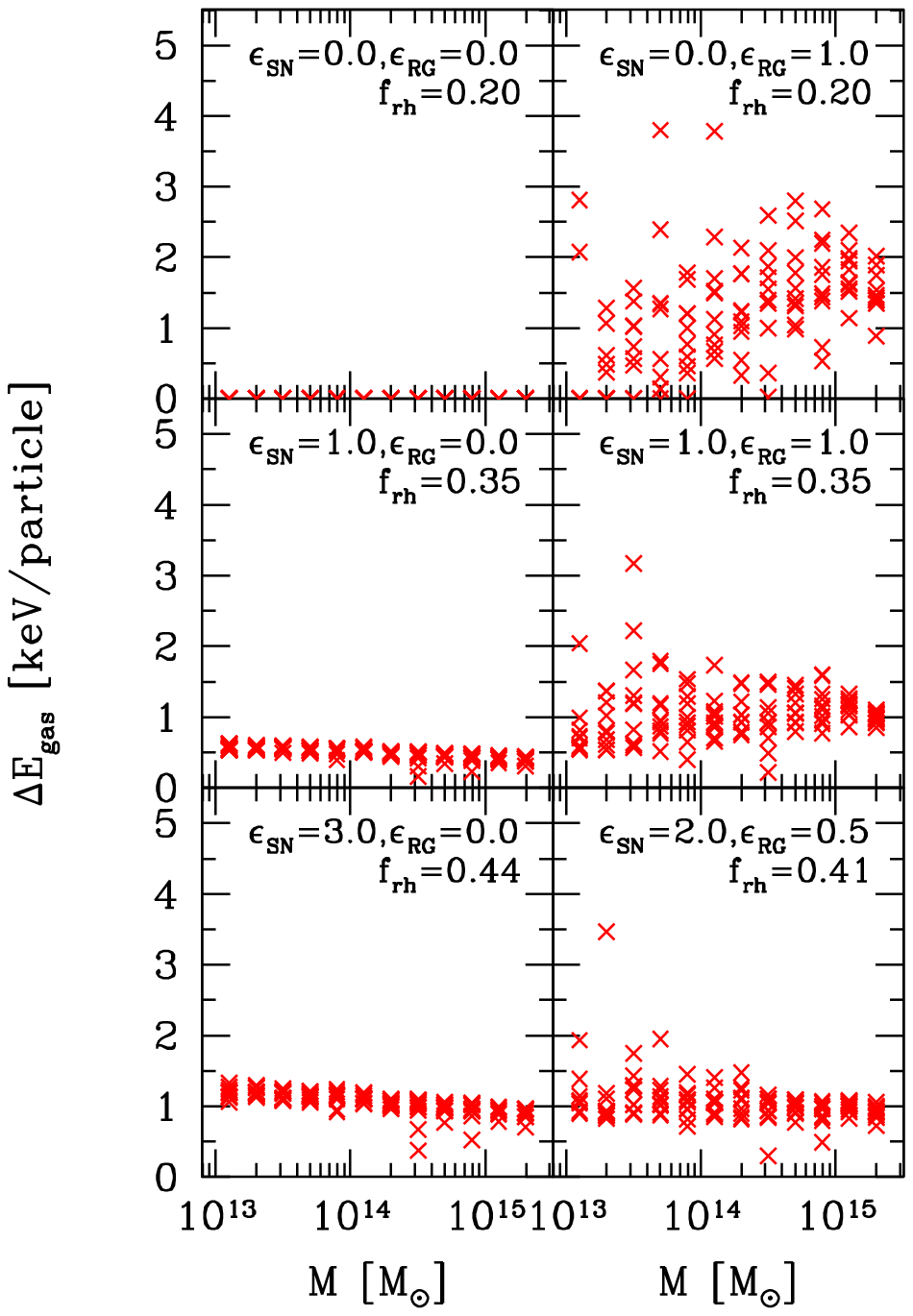}
\caption{Scatter plots of the excess energy per gas particle of clusters
  at $z=0$. Our model predictions are plotted in crosses.
    \label{fig:ExcessEnergy}}
\end{figure}

Finally the excess energy per gas particle, $\Delta E_{\mathrm{gas}}$,
is plotted in figure~\ref{fig:ExcessEnergy} against the halo mass at
$z=0$. As expected, all the successful models (middle and bottom
panels) have $\Delta E_{\mathrm{gas}} \sim 1$keV/particle.  As we have
seen in other plots, the jet from radio galaxies increases the
dispersion.  Note that the excess energy is defined as per one hot gas
particle; the hot gas mass fraction in the top-right panel is merely
30 percent and this is why $\Delta E_{\mathrm{gas}}$ ranges up to
$\sim 2$keV/particle although the total heating energy is less than
the other models (middle and bottom panels).

%%%%%%%%%%%%%%%%%%%%%%%%%%%%%%%%%%%%%%%%%%%%%%%%%%%%%%%%%%%%%%%%%%%%%%
%%%%%%%%%%%%%%%%%%%%%%%%%%%%%%%%%%%%%%%%%%%%%%%%%%%%%%%%%%%%%%%%%%%%%%

As we mentioned in Introduction, our current approach corresponds to
the improved modeling of the earlier work by \citet{wu00}. Thus it is
relevant to compare the results here.  They concluded that the excess
energy of $\sim (1\mbox{--}3)$keV/particle is required to reproduce
the observed luminosity -- temperature relation. This is in good
agreement with others as well as our results
(figure~\ref{fig:ExcessEnergy}).  They also argued that this amount of
excess energy is unlikely to be provided by supernova feedback except
in a highly contrived model of galaxy formation, and suggested the
AGNs as a potentially important heating source. In this respect, the
normalization of our model parameter, $\epsilon_{\scriptscriptstyle\rm
  SN}$, may be a bit optimistic; they assume the average kinetic
energy from supernova of $4\times 10^{50}
\epsilon_{\scriptscriptstyle\rm Wu, SN}$erg with their efficiency
parameter $\epsilon_{\scriptscriptstyle\rm Wu, SN} \sim 0.3$, while we
adopted $10^{51}\epsilon_{\scriptscriptstyle\rm SN}$erg.  Since they
did not show plots of the statistical properties of their resulting
clusters as we did in the present section, further quantitative
comparison is difficult, but the above overall qualitative conclusions
seem to be in reasonable agreement with each other.

%%%%%%%%%%%%%%%%%%%%%%%%%%%%%%%%%%%%%%%%%%%%%%%%%%%%%%%%%%%%%%%%%%%%%%
\section{Model of higher star formation activities at $z>7$}
\label{sec:highsn}
%%%%%%%%%%%%%%%%%%%%%%%%%%%%%%%%%%%%%%%%%%%%%%%%%%%%%%%%%%%%%%%%%%%%%%

One reasonable interpretation of the recent \textit{WMAP} result on
the early reionization at $z=17\pm 5$ \citep{Spergel03} is that the
star formation rate was much higher and/or the energy injection from
supernovae was stronger; it has been argued  that the first objects at
high redshifts may be preferentially very massive ($\sim
100\;M_{\odot}$), and that their birth rate may be higher than those
in the range $1\mbox{--}2\;M_{\odot}$ (e.g., \cite{bromm99};
\cite{abel00}; \cite{NU01}). Such first objects may turn into
hypernovae with the explosion energy $\gtrsim 10^{52}\;\mathrm{erg}$.
In order to mimic such \textit{plausible} scenarios, we consider a
model with the bimodal supernova feedback efficiency:
%%%%%%%%%%%%%%%%%%%%%%%%%%%%%%%%%%%%%%%%%%%%%%%%%%%%%%%%%%%%%%%%%%%%
\begin{equation}
  \epsilon_{\scriptscriptstyle\mathrm{SN}}=
\begin{cases}
  {\epsilon_{\scriptscriptstyle\mathrm{SN,L}} & ($z < 7$) \cr
    \epsilon_{\scriptscriptstyle\mathrm{SN,H}} & ($z \geq 7$) }
\end{cases} .
\end{equation}
%%%%%%%%%%%%%%%%%%%%%%%%%%%%%%%%%%%%%%%%%%%%%%%%%%%%%%%%%%%%%%%%%%%%
For definiteness we fix that
$\epsilon_{\scriptscriptstyle\mathrm{SN,L}}=0.3$, corresponding to the
30 percent conversion efficiency of the supernova energy to the ICM
thermalization. Since the efficiency at high $z$ is much more
uncertain, we explore the four cases;
$\epsilon_{\scriptscriptstyle\mathrm{SN,H}}=1.0$, 2.0, 5.0, and 10.0.
We neglect the jet heating, i.e.,
$\epsilon_{\scriptscriptstyle\mathrm{RG}}=0$, and adopt that
$f_{\mathrm{rh}}=0.26$ [equation~(\ref{eq:frh}) with
$\epsilon_{\scriptscriptstyle\mathrm{SN}} = 0.3$] at all redshifts.
%%%%%%%%%%%%%%%%%%%%%%%%%%%%%%%%%%%%%%%%%%%%%%%%%%%%%%%%%%%%%%%%%%%%%%
\begin{figure}[tbh]
  \centering \FigureFile(80mm,80mm){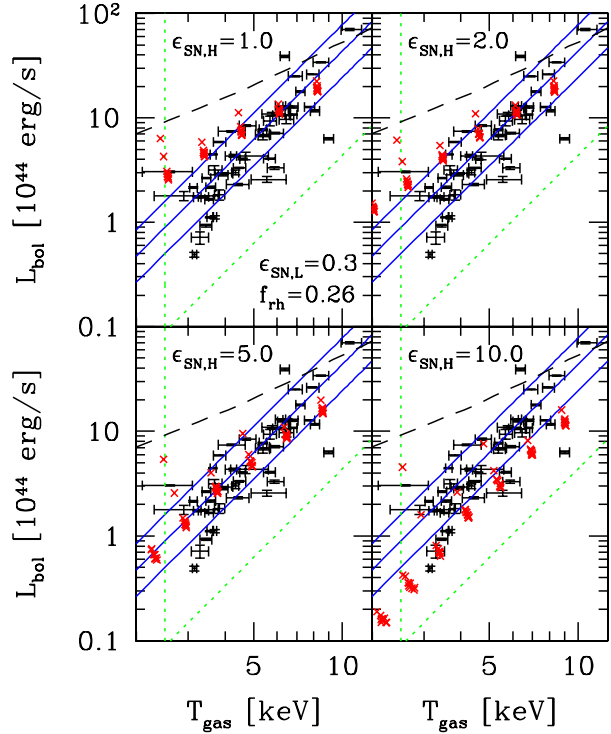}
  \caption{Same as figure~\ref{fig:lt} but for models with enhanced star
    formation activity for $z>7$.  We assume that
    $\epsilon_{\scriptscriptstyle\mathrm{RG}}=0$ and
    $f_{\mathrm{rh}}=0.26$ at all redshifts.  \label{fig:bimlt}}
\end{figure}
%%%%%%%%%%%%%%%%%%%%%%%%%%%%%%%%%%%%%%%%%%%%%%%%%%%%%%%%%%%%%%%%%%%%%%

Figure~\ref{fig:bimlt} shows that the luminosity-temperature relation
of the model clusters for
$\epsilon_{\scriptscriptstyle\mathrm{SN,H}}\sim 5.0$ (lower left
panel) reproduces the observation reasonably well.  This may
correspond to an original idea of \textit{preheating} by \citet{eh91}
and \citet{kaiser91}, and thus the result is fairly insensitive to
non-gravitational heating at lower redshifts.

\begin{figure}[tbh]
  \centering \FigureFile(80mm,80mm){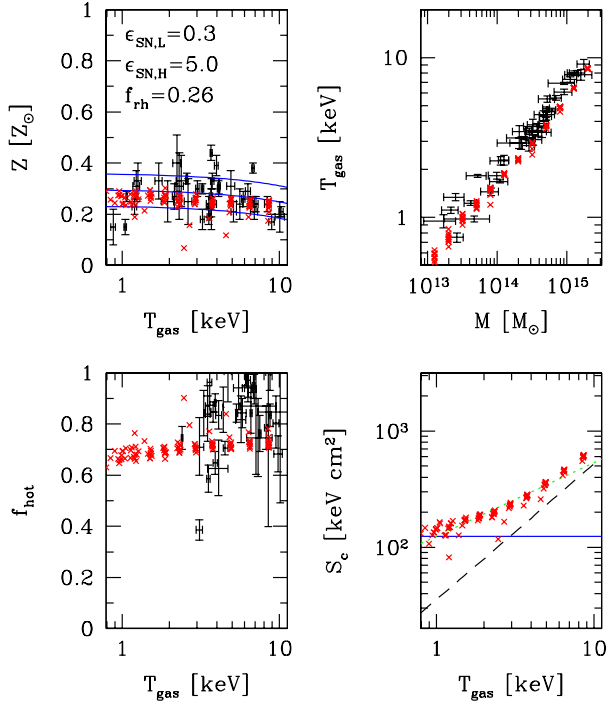}
  \caption{Scatter plots of the metallicity-temperature relation, the
    mass-temperature relation, the hot gas fraction, and the core
    `entropy' for models with enhanced star formation activity for
    $z>7$.  Our model predictions are plotted in crosses. We assume
    that $\epsilon_{\scriptscriptstyle\mathrm{SN,L}}=0.3$ and
    $\epsilon_{\scriptscriptstyle\mathrm{SN,H}}=5.0$, while the other
    parameters have the same values as in figure~\ref{fig:bimlt}.
    \label{fig:bimother}}
\end{figure}

The other relations are shown in figure~\ref{fig:bimother} for
$\epsilon_{\scriptscriptstyle\mathrm{SN,L}}=0.3$ and
$\epsilon_{\scriptscriptstyle\mathrm{SN,H}}=5.0$.  Since we did not
attempt to tune the parameter sets, we would say that the agreement is
satisfactory; the inclusion of the jet heating by radio galaxies at
low redshifts would certainly improve the agreement.  These results
are impressive because we do not have to consider exceedingly strong
supernova feedback at low redshifts if the star formation activity at
high redshifts is sufficiently enhanced.

%%%%%%%%%%%%%%%%%%%%%%%%%%%%%%%%%%%%%%%%%%%%%%%%%%%%%%%%%%%%%%%%%%%%%%
%%%%%%%%%%%%%%%%%%%%%%%%%%%%%%%%%%%%%%%%%%%%%%%%%%%%%%%%%%%%%%%%%%%%%%

%%%%%%%%%%%%%%%%%%%%%%%%%%%%%%%%%%%%%%%%%%%%%%%%%%%%%%%%%%%%%%%%%%%%%%
\section{Summary and conclusions}
%%%%%%%%%%%%%%%%%%%%%%%%%%%%%%%%%%%%%%%%%%%%%%%%%%%%%%%%%%%%%%%%%%%%%%

We have developed a semi-analytical approach to trace the thermal
history of galaxy clusters based on the Monte-Carlo modeling of
merging trees of dark matter halos. Under the assumption of
hydrostatic equilibrium and the isothermal gas profiles, we have
incorporated the metallicity evolution, the metallicity-dependent
cooling of gas, the supernova energy feedback, and heating due to the
jet of radio galaxies in a consistent manner. The latter
non-gravitational heating processes were characterized by two
dimensionless parameters, $\epsilon_{\scriptscriptstyle\mathrm{SN}}$
and $\epsilon_{\scriptscriptstyle\mathrm{RG}}$, and we explored
several statistical properties of galaxy clusters over a wide range of
the parameter space.

As has been known for a while, we confirmed that a fiducial model of
supernova feedback alone, i.e.,
$\epsilon_{\scriptscriptstyle\mathrm{SN}}<1$, does not reproduce the
observed luminosity -- temperature relation of clusters.  A reasonable
agreement can be achieved by enhancing non-gravitational heating in
two different ways; i) considering additional heating due to the jet
of some class of AGNs, notably Type II of the Fanaroff-Riley radio
galaxies, and ii) adopting somewhat higher star formation efficiency
and/or supernova energy feedback. The former possibility was first
examined seriously by \citet{IS01}, and the current study basically
confirmed their conclusion using a significantly improved methodology
to trace the thermal history of ICM.  The latter idea is particularly
interesting in the light of the recent \textit{WMAP} finding of the
earlier reionization epoch of the universe than previously thought.
By increasing the feedback efficiency at high redshifts
($\epsilon_{\scriptscriptstyle\mathrm{SN}}\sim 5$ at $z>7$, for
instance), most of the model predictions of the simulated clusters can
be brought into agreement with the observational data.

So far we discussed several properties of clusters at $z=0$, and did
not examine their evolution. Considering the success of the enhanced
star formation activity model, it is important to combine the ICM
heating model with the cosmic star formation history. In doing that,
we will have tighter constraints on the parameter space of
$\epsilon_{\scriptscriptstyle\mathrm{SN}}$ and
$\epsilon_{\scriptscriptstyle\mathrm{RG}}$, and may have a link to the
physical reasonable scenario beyond a general but parameterized
modeling like our current approach. A future sample of clusters at
high $z$ selected by the Sunyaev-Zel'dovich effect \citep{sz72} may
provide another complementary piece of information on the thermal
evolution of the ICM. We hope to report results on these important
issues elsewhere in due course.

%%%%%%%%%%%%%%%%%%%%%%%%%%%%%%%%%%%%%%%%%%%%%%%%%%%%%%%%%%%%%%%%%%%%%%%%%%%
\bigskip

This research was supported in part by the Grant-in-Aid for Scientific
Research of JSPS (12640231, 14102004, 14740133, 15740157). Numerical
computation was performed using computer facilities at the University
of Tokyo supported by the Special Coordination Fund for Promoting
Science and Technology, Ministry of Education, Culture, Sport, Science
and Technology.

\appendix
%%%%%%%%%%%%%%%%%%%%%%%%%%%%%%%%%%%%%%%%%%%%%%%%%%%%%%%%%%%%%%%%%%%%%%
\section*{Tracing thermal evolution of ICM in merging trees of dark
  matter halos}
%%%%%%%%%%%%%%%%%%%%%%%%%%%%%%%%%%%%%%%%%%%%%%%%%%%%%%%%%%%%%%%%%%%%%%

In the present paper, we trace merger trees from $z=0$ to $z=30$.  In
this case, however, the number of progenitors in one merging tree
becomes progressively larger at higher redshifts, and it is not
practically feasible. Thus we decide to prepare two different sets of
merger trees at low redshifts ($z=0$ to $z=7$) and at high redshifts
($z=7$ to $z=30$).

We first construct $N_{\rm ens}$ (=10) independent sets of merger tree
realizations for each value of 12 different masses of $M_{\rm root}$
starting from $z_{\rm min}=0$ to $z_{\mathrm{max}}=7$ in $N_{\rm
  step}=750$ timesteps.  We also construct separately 20 sets of
merger tree realizations starting from $z_{\rm min}=7$ to
$z_{\mathrm{max}}=30$ in $N_{\rm step}=500$ timesteps.  We use the
number of timesteps in logarithmically equal redshift interval,
$N_{\rm step}=750$ for the low $z$ sets and $N_{\rm step}=500$ for the
high $z$ sets.  The masses of the root halo $M_{\mathrm{root}}$ are
equal to $10^{13+0.2(i-0.5)}\;M_{\odot}\ (i=1, \dots, 12)$ at $z=0$
for the low $z$ sets, and $2.0\times 10^{7+j}\;M_{\odot}\ (j=1, \dots,
6)$ at $z=7$ for the high $z$ sets.  The minimal mass of progenitors
resolved in the merger tree $M_{\rm res}$ is equal to $M(T_{\rm
  vir}=10^4\;{\rm K})$, which corresponds to a mass of halos whose
virial temperature is $10^4\;{\rm K}$ at each redshift. The parameters
concerning the merging tree of halos is summarized in
table~\ref{tab:treeparam}.
%%%%%%%%%%%%%%%%%%%%%%%%%%%%%%%%%%%%%%%%%%%%%%%%%%%%%%%%%%%%%%%%%%%%%%
\begin{table}[thb]
  \caption{Parameters for  merger trees of dark halos.
    \label{tab:treeparam}}
  \begin{center}
    \begin{tabular}{cccl}\hline\hline
      \multicolumn{1}{c}{symbol} &
      \multicolumn{2}{c}{adopted value} &
      \multicolumn{1}{c}{physical meaning} \\
      \multicolumn{1}{c}{} &
      \multicolumn{1}{c}{low redshift} &
      \multicolumn{1}{c}{high redshift} &
      \multicolumn{1}{c}{} \\\hline
      $M_{\rm root}$ & $10^{13+0.2(i-0.5)}\;M_{\odot}$
      & $2.0\times 10^{7+j}\;M_{\odot}$ &mass of halo
      at $z=z_{\mathrm{min}}$ \\
      & $(i=1, \dots, 12)$ & $(j=1, \dots, 6)$ & \\
      $M_{\rm res}$ & \multicolumn{2}{c}{$M(T_{\rm vir}=10^4\;{\rm
      K})$} & minimal mass of progenitors resolved in each merger tree\\
      $N_{\rm step}$ & 750 & 500
      & number of  redshift bins (logarithmically equal interval)\\
      $N_{\rm ens}$ & 10 & 20 & number of realizations of the merger tree\\
      $z_{\rm min}$ & 0 & 7 & minimum redshift of the merger tree \\
      $z_{\rm max}$ & 7 & 30 & maximum redshift of the merger tree \\ \hline
    \end{tabular}
  \end{center}
\end{table}
%%%%%%%%%%%%%%%%%%%%%%%%%%%%%%%%%%%%%%%%%%%%%%%%%%%%%%%%%%%%%%%%%%%%%%

Assuming that baryons at $z=30$ have their gas temperature equal to
the virial temperature of their individual host halo, both the halo
merging history and the thermal evolution of gas are followed up to
$z=7$.  At that redshift, each halo is assigned to another realization
of the halo in the separate redshift merger trees using the
interpolation of the mass in finding its counterpart. Then all the
properties of the halo at high redshift tree is now transferred to its
counterpart at the low redshift trees. Then we continue to follow its
evolution until $z=0$ tracing the new merging tree.

We checked the validity of the above interpolation method explicitly
by constructing the merger trees from $z=0$ to $z=10$, and tracing the
evolution using each merger tree. The comparison with the results
based on statistically connecting two separate merger trees from $z=0$
to $z=7$ and from $z=7$ to $z=10$ indicates that the properties of
clusters at $z=0$ are almost indistinguishable.

\clearpage
%%%%%%%%%%%%%%%%%%%%%%%%%%%%%%%%%%%%%%%%%%%%%%%%%%%%%%%%%%%%%%%%%%%%%%%%%%%

%%%%%%%%%%%%%%%%%%%%%%%%%%%%%%%%%%%%%%%%%%%%%%%%%%%%%%%%%%%%%%%%%%%%%%%%%%%%%

\end{document}